\documentclass{aa}  

\usepackage{graphicx}
\usepackage{txfonts}
\usepackage[dvipsnames]{xcolor}
\usepackage{multirow}
\usepackage{array}
\usepackage{float}
\usepackage[normalem]{ulem}
\usepackage{pifont}
\usepackage{amsmath}

\begin{document} 

   \title{Gaia-Sausage-Enceladus star formation history as revealed by detailed elemental abundances}

   \subtitle{An archival study using SAGA data}

   \author{H. Ernandes\thanks{\email{heitor.ernandes@geol.lu.se}}
          \inst{1}
          \and
          D. Feuillet\inst{1,2}
          \and
          S. Feltzing\inst{1}
          \and
          Á. Skúladóttir\inst{3}
          }

   \institute{Lund Observatory, Department of Geology, S\"olvegatan 12, SE-223 62 Lund, Sweden
   \and
   Observational Astrophysics, Department of Physics and Astronomy, Uppsala University, Box 516, SE-751 20 Uppsala, Sweden
   \and
   Dipartimento di Fisica e Astronomia, Universitá degli Studi di Firenze, Via G. Sansone 1, I-50019 Sesto Fiorentino, Italy
              }

   \date{Received XXXX; accepted YYYY}

  \abstract
   {The Gaia-Sausage-Enceladus merger was a major event in the history of the Milky Way. Debris from this merger has been extensively studied with full kinematic data from the Gaia mission. Understanding the star formation history of the progenitor galaxy aids in our understanding of the evolution of the Milky Way and galaxy formation in general.
   }
   {We aim to constrain the star formation history of the Gaia-Sausage-Enceladus progenitor galaxy using elemental abundances of member stars. Previous studies on Milky Way satellite dwarf galaxies show that key elemental abundance patterns, which probe different nucleosynthetic channels, reflect the host galaxy's star formation history.
   }
   {We gather Mg, Fe, Ba, and Eu abundance measurements for Gaia-Sausage-Enceladus stars from the SAGA database. Gaia-Sausage-Enceladus members are selected kinematically. Inspired by previous studies, we use [Fe/Mg], [Ba/Mg], [Eu/Mg], and [Eu/Ba], as a function of [Fe/H] to constrain the star formation history of Gaia-Sausage-Enceladus.  We use the known star formation histories and elemental abundance patterns of the Sculptor and Fornax dwarf spheroidal galaxies as comparison.
   }
   {The elemental abundance ratios of [Fe/Mg], [Ba/Mg], [Eu/Mg], and [Eu/Ba] all increase with [Fe/H] in Gaia-Sausage-Enceladus. The [Eu/Mg] begins to increase at [Fe/H] $\sim -2.0$ and continues steadily, contrasting with the Sculptor dSph galaxy. The [Eu/Ba] increases and remains high across the [Fe/H] range, contrasting with that of the Sculptor dSph galaxy and deviating from the Fornax dSph galaxy at high [Fe/H]. The [Ba/Mg] is higher than those of the Sculptor dSph galaxy at the lowest [Fe/H] and gradually increases, similar to the Fornax dSph galaxy.
   We constrain three main properties of the Gaia-Sausage-Enceladus star formation history: 1) star formation started gradually, 2) it extended for over 2 Gyr, and 3) it was quenched around [Fe/H] of $-0.5$, likely when it fell into the Milky Way. }
   {We show that the elemental abundance ratios [Fe/Mg], [Ba/Mg], [Eu/Mg], and [Eu/Ba] can be used to trace the star formation history of a disrupted galaxy when these measurements are available over an [Fe/H] range that is representative of the progenitor galaxy's stellar population.
   }


   \maketitle


\section{Introduction}

Recently, using astrometric data from ESA's Gaia satellite, the presence of a large galaxy that merged with the Milky Way about 10 billion years ago was confirmed and dubbed the Gaia-Sausage-Enceladus population \citep[e.g.,][]{Helmi18,Belokurov18}. Indications of such a population had been noted in the literature as early as \citet{Nissen10}. 
With larger spectroscopic and astrometric datasets available, several studies have now looked into how to select the stars that came with that galaxy using kinematics and/or chemical composition \citep[see][and references therein]{Belokurov18, Feuillet21,Carrillo23}. 

To first order, the surface composition of stars reflects the gas composition from which they formed. The relative abundance of individual elements in the gas is strongly dependent on the history of the host galaxy as different elements are deposited into the interstellar medium on different timescales. The $\alpha$-elements (e.g. O and Mg) are all (mainly) released from ccSN \citep{WW95}. This process will enrich the interstellar medium on short timescales ($10^{7-8}$~yr), tracking the star formation. The iron-peak elements on the other hand, are mainly produced by type Ia SN (SNIa), which happens on longer timescales ($10^{8-9}$~yr).

Elements heavier than Zn ($Z>30$) are created through neutron-capture nucleosynthesis. This process can be divided into subcategories depending on the flux of neutrons required to create them. These subcategories are the slow~($s$), intermediate~($i$), and rapid~($r$) processes. The $s$-process takes place when the neutron flux is lower than the time of the $\beta$-decay (with which the nucleus moves towards higher proton numbers). Thus it is feasible for this process to occur in Asymptotic Giant Branch (AGB) stars on long timescales ($10^{8-10}$ yr, see review by \cite{Kappeler11} and references therein). The $s$-process is the main nucleosynthetic site of elements such as Ba. The $r$-process on the other hand requires a faster neutron capture. The exact site for this process is still a matter of debate. The $r$-process elements, such as Eu, likely have more than one source and have been found to contribute to the chemical enrichment of a stellar population both on long and short time-scales~\citep[][]{Skuladottir20}. 

The abundance of a given elements over time in the interstellar medium of a galaxy will depend on the timescale of the contributing stellar source(s) and how many stars have been formed. The star formation history of a given (dwarf) galaxy will, therefore, leave a unique imprint in the stellar elemental abundances. Several studies show that the Gaia-Sausage-Enceladus galaxy has an elemental abundance pattern for the $\alpha$-elements that differs from that of the rest of the Milky Way halo and the metal-weak thick disk at the same metallicities \citep[see, e.g.][]{Nissen10, Nissen24,Helmi18}. 
This provides strong evidence that stars belonging to the Gaia-Sausage-Enceladus population formed in a galaxy outside the Milky Way.

The $\alpha$-element pattern and overall metallicity distribution function of the Gaia-Sausage-Enceladus have been well characterised \citep[e.g.][]{Helmi18, Feuillet20, Feuillet21, Naidu20}.  The metallicity range of Gaia-Sausage-Enceladus in the literature is generally found to span $\rm-2.0 < [Fe/H] < -0.5$, although the bounds vary by up to $\sim 0.3$ dex depending on the exact selection scheme used \citep[e.g.][]{Naidu20,Feuillet20,Hasselquist21,Bonifacio21}.
Studies of elemental abundance trends beyond the $\alpha$-elements are few and, most importantly,  many do not cover the full metallicity range \citep{Aguado21, Matsuno22, Monty20}. 
 The sample of \citet{Matsuno21} uses data from GALAH DR3 to select a sample of Gaia-Sausage-Enceladus members that spans $\rm -1.7 < [Fe/H] < -0.5$. However, the number of stars with reliable Eu measurements is not sufficient to fully characterise the abundance trends at the low metallicity end.
This makes it difficult to exploit the elemental abundances from different nucleosynthetic channels in order to constrain the full star formation history of the Gaia-Sausage-Enceladus galaxy, particularly because the metal-poor end is very important to constrain details of the galaxy's early star formation. 
In this work, we build on \cite{Skuladottir20} to do just that.

\cite{Skuladottir20} use a combination of star formation histories for different dwarf spheroidal (dSph) galaxies together with elemental abundance patterns to put constraints on the timescales of the $r$-process sites. Inspired by this result, we find that if we have elemental abundances measured for stars in Gaia-Sausage-Enceladus, we can add the known nucleosynthesis timescales to interpret the features we observed in the abundance trends as constraints to its star formation history.

In this paper, we take advantage of the availability of a large body of data from the literature as compiled in the SAGA database \citep{SAGA1}, which provides us with data for halo (and disk) stars. Using the selection criteria described in Sec. \ref{sect:analysis} for selecting the stars, we proceed to trace the star formation history of the Gaia-Sausage-Enceladus galaxy using these data. We use star formation histories and elemental abundance patterns from the literature of the dSph galaxies Fornax and Sculptor as comparisons to aid our interpretation. Critically, the abundance measurements of Mg, Ba, and Eu over a large range in [Fe/H] allow us to constrain the star formation history of Gaia-Sausage-Enceladus in detail.

\section{Data}
\label{sect:data}

\begin{figure}
    \centering
    \includegraphics[width=0.48\textwidth]{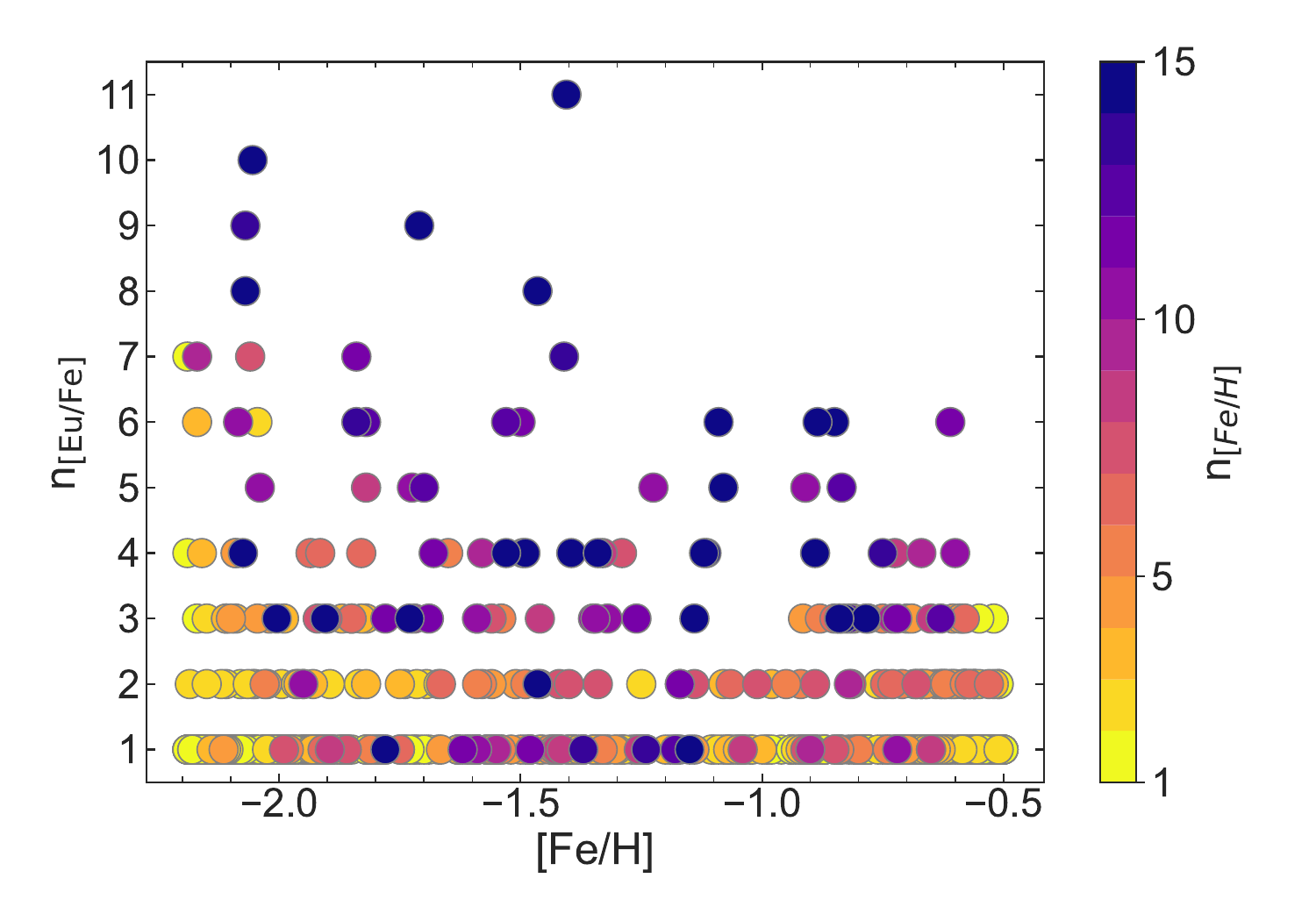}
    \caption{Number of [Eu/Fe] measurements (n[Eu/Fe]) for the stars in our SAGA catalogue as a function of the median [Fe/H]. Colour coding shows the number of available [Fe/H] measurements (n[Fe/H]). 
    }
    \label{multipleValues}
\end{figure}

\begin{figure}
    \centering
    \includegraphics[width=0.48\textwidth]{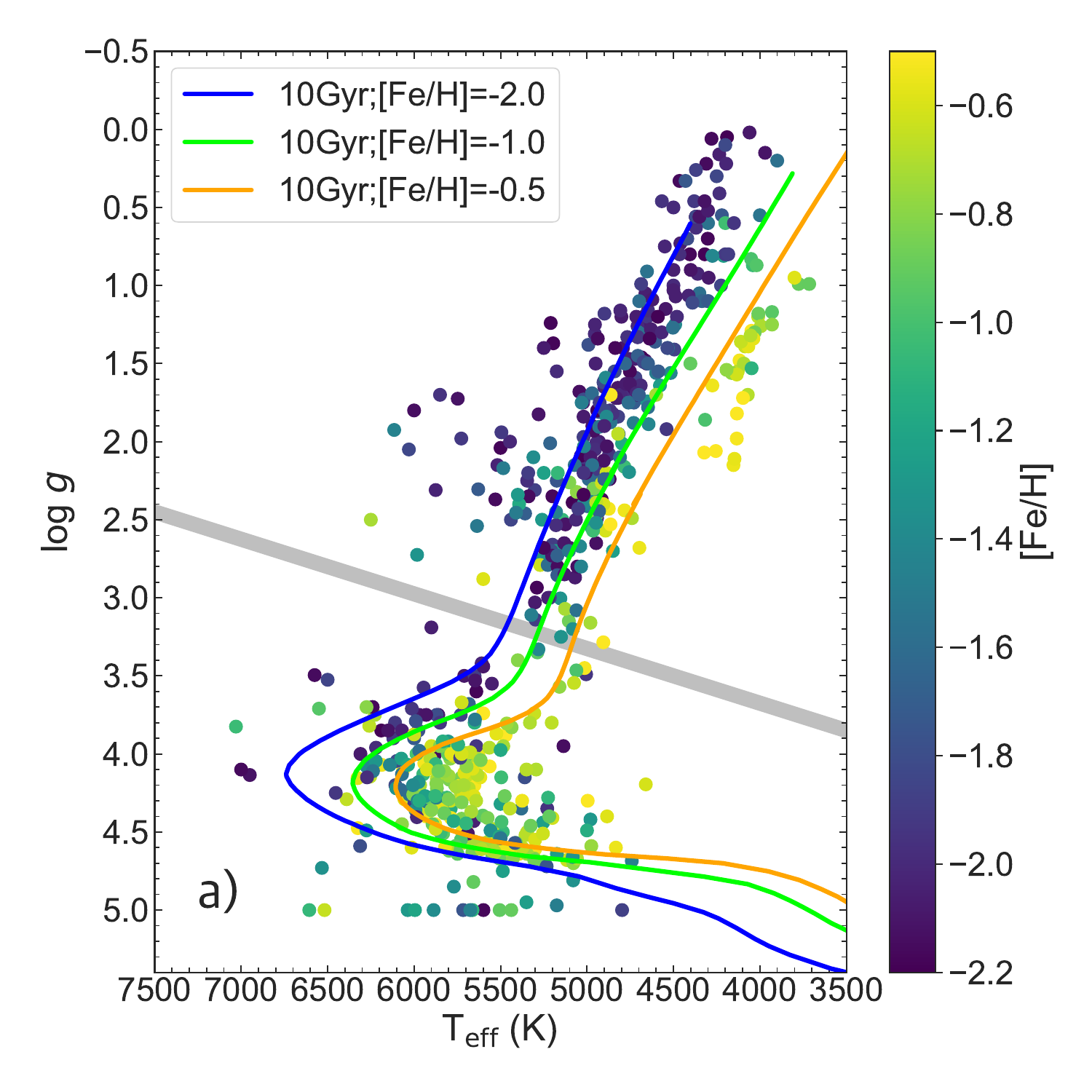}
    \caption{Properties of the stars in our final sample with SAGA and Gaia data, with three 10 Gyr PARSEC isochrones overlaid, metallicity as indicated in the legends: Kiel diagram, colour-coded by the median [Fe/H]. The shaded grey line represents the visual definition we used in Sec. \ref{sect:analysis} to divided the sample into dwarf and giant stars.
    }
    \label{fullCMDHR}
\end{figure}

\begin{table*}
\centering
    \caption{Gaia ID, SAGA ID, median SAGA stellar parameters, elemental abundances, photometric $T_{\rm eff}$ (only for dwarf stars), median age with 16 and 84 percentile, and $r$-II flag for stars selected according to \cite{Feuillet21} scheme.  The full table is available in electronic format.}
    \renewcommand{\arraystretch}{1.2} 
    \resizebox{\textwidth}{!}{
    \begin{tabular}{llcccccccccccc}
    \hline
    \hline
  Gaia DR3 ID & SAGA ID & $T_{\rm eff}$ & log$g$ &  [Fe/H] & [Mg/Fe] & [Eu/Fe] & [Ba/Fe] & $T_{\rm eff}$ 
    & a16 & Age & a84 & $r$-II \\
   &&[K]&&&&&&[K]&[Gyr]&[Gyr]&[Gyr]& \\
   \cline{3-13}
     &&SAGA&&&&&&Photo&&Median&& flag  \\
    \hline
    1458016709798909952 & G165-39 & 6220 & 3.96 & $-$1.99 & +0.22 & +0.48 &$-$0.04 & 6405 & 11.7 &  13.0 & 14.3 & No \\  
   689795494613436160 & G9-36 & 5788 & 4.35 & $-$1.14 & +0.29 &+0.49 &+0.11 & 5872 & --- & --- & --- & No \\
   2077681190974623744 & KIC6611219 & 4652 & 1.74 & $-$1.32 & +0.26 & +0.92 & +0.20 & --- & --- & --- & --- & Yes \\
   \multicolumn{13}{c}{$\cdots$} \\
    \hline
    \hline
    \label{tab:data}
    \end{tabular}
    }
\end{table*}

\subsection{Chemical abundances in SAGA}

The goal of this paper is to derive the Gaia-Sausage-Enceladus star formation history using key elemental abundance trends. To do this we need to include stars that cover the full range of [Fe/H].
We are in particular interested in the neutron capture elements and the abundance trends at low metallicity, which reflects the early stages of star formation. To our knowledge, \citet{Matsuno21} is the only study of $r$- and $s$-process elements in Gaia-Sausage-Enceladus with more than 30 stars, covering $\rm-1.7<[Fe/H]<-0.4$. However, the GALAH DR3 Eu measurements that they use are flagged as lower quality at the lowest metallicity end of the Gaia-Sausage-Enceladus galaxy. As the SAGA database was historically a catalogue for studies of metal-poor stars, it is a good resource for elemental abundance measurements of stars in the metal-poor tail of Gaia-Sausage-Enceladus.

In order to gather the largest sample possible with measurements of neutron capture elements across a large range of metallicity, we chose to utilise a compilation of data from the literature, the SAGA database. 
The SAGA database \citep{SAGA1} gathers elemental abundance measurements from the literature provided by papers in the CDS\footnote{Centre de Données astronomiques de Strasbourg} database. The SAGA database was originally designed to provide a catalogue of metal-poor stars. However, recently studies of more metal-rich stars have been included. The database is categorised into catalogues of several dwarf galaxy and the Milky Way. We selected stars from the latter catalogue as described below. The version of the database used here was last updated on April 10, 2023.

Using the SAGA plotting interface\footnote{http://sagadatabase.jp/}, we selected stars within the iron (Fe) abundance range we expect for Gaia-Sausage-Enceladus stars \citep[e.g.][]{Feuillet20} and extended to slightly lower [Fe/H] than seen in many large surveys that focus on the Milky Way disk. We selected stars meeting the following criteria:
\begin{itemize}
    \item[i)] $-2.2 <$ [Fe/H] $< -0.5$,  
    \item[ii)] measurements available for both [Eu/Fe] and [Fe/H].
\end{itemize}
This selection resulted in a sample of 804 stars, however, we note that after searching for Gaia DR3 IDs for this sample, we find that not all of these stars are unique, see Sec.~\ref{sec:Gaia} for details. For these stars, we added [Mg/Fe] and [Ba/Fe] measurements from the SAGA database when available. We remove from our sample  Ba-rich stars, which are stars which have obtained large amounts of Ba (and other elements such as C) via binary transfer, and therefore do not trace the chemical evolution of the host system, see App.~\ref{app:Bastars} for details.

The resulting catalogue provides for every star a unique row for each literature source contributing a reported [Fe/H], [Eu/Fe], [Mg/Fe], and [Ba/Fe] as shown in table \ref{tab:data}. This means that every star can have multiple entries  (see App. \ref{sect:app_refs} and table \ref{tab:refs-diane} for all references), and each elemental abundance measurement may come from a different study. Some literature studies only derive elemental abundances and use stellar parameters, including [Fe/H], from other sources. Therefore, the stellar parameters of effective temperature ($T_{\rm eff}$), surface gravity ($\log g$), and [Fe/H] may be neither unique nor homogeneous for each reported elemental abundance of a single star. Fig.~\ref{multipleValues} illustrates the number of [Eu/Fe] and [Fe/H] measurements available for stars in our catalogue.

The potential for multiple reported elemental abundances for a single star means that we needed to make choices about how to combine the elemental abundances available. When multiple abundance values are reported for a single element in the same star, we use the median of all values. The systematic uncertainties from each literature study were not considered in this analysis. In section~\ref{error}, we discuss the uncertainty in our results due to element abundance dispersion between multiple studies and how this is indicated in later figures.

Fig.~\ref{fullCMDHR} shows the Kiel diagram for the final sample of 654 stars from the SAGA database that also have Gaia parameters (see discussion in Section~\ref{sec:Gaia}). As was done with the elemental abundances, we take the median $T_{\rm eff}$ and log$g$ values of all reported literature values for a given star. The colour of each point indicates the median [Fe/H] of each star.
We note the metal-rich giant branch stars are cooler than the majority of stars in the sample. These stars, which all come from \citet{tautvaisien21}, are considerably more metal-rich ($\rm [Fe/H]>-0.7$) than the other giant branch stars, therefore the separation is expected. The metallicity distribution of our final sample is shown in grey in Fig.~\ref{MDF}.

\subsection{Astrometric data and kinematics}
\label{sec:Gaia}

We use Gaia astrometric parameters to determine the kinematics of the stars in our sample. To do this, we first query for the Gaia DR3 IDs using the capabilities of the \textsc{astroquery.simbad} \citep{astroquery19,Simbad00} package in Python. In this process, we find that 13 stars from \citet{Li22}, which are in our SAGA sample, are not in the SIMBAD database under the ID provided by SAGA. Through personal communication with Haining Li, we received the Gaia DR3 ID for these 13 stars. This results in a sample of 680 unique stars from SAGA with Gaia DR3 IDs. 

We then query the Gaia archive\footnote{https://gea.esac.esa.int/archive/} for the Gaia DR3 parameters \citep{GaiaDR32021}, finding that 654 stars have Gaia astrometric and Radial Velocity Spectrometer (RVS) data available.
We calculate the full kinematics and orbital parameters of the stars in the sample using the Python packages \textsc{astropy} \citep{astropy13, astropy18, astropy22} and \textsc{galpy} \citep{galpy15}. When determining the actions, we assume the \textsc{actionAngleStaeckel} approximation \citep{Binney2012, Bovy2013} using a delta of 0.4. We calculate actions and kinematics using both the \textsc{MWPotential14} \citep{Bovy2013} and \citet{McMillan2017} models for the Milky Way potential as this is important when comparing different Gaia-Sausage-Enceladus selection schemes reported in the literature. We adopted the geometric distances provided in \cite{Bailer-Jones21}.

We use the spectroscopic radial velocity measurement from the SAGA database when available. In the case that SAGA provided multiple RV entries for the same target, we decided to adopt the value from the newest reference as the deviation between different literature values for the radial velocity is small. When radial velocity is not available in SAGA, we used the Gaia RVS radial velocity.
The kinematics are used to select the Gaia-Sausage-Enceladus samples, see Sect.\,\ref{GSEselect-sect}. 

\section{Analysis}
\label{sect:analysis}

\subsection{Selecting the Gaia-Sausage-Enceladus stars}\label{GSEselect-sect}

\begin{figure}
    \centering
    \includegraphics[width=0.45\textwidth]{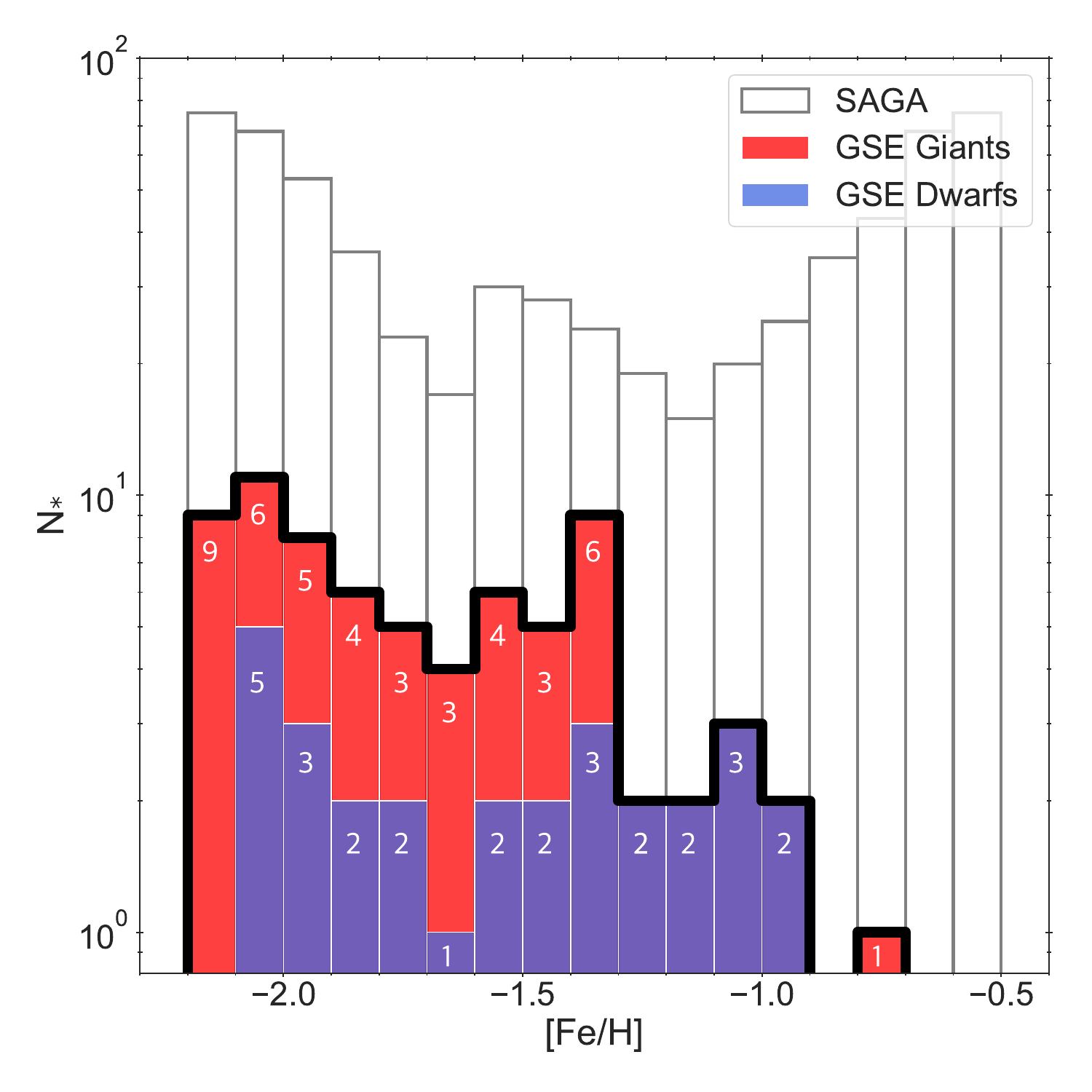}
    \caption{
    Metallicity distribution for the stars in our final catalogue with SAGA and Gaia data (grey outline). The stars in our primary Gaia-Sausage-Enceladus sample are also shown, dwarf stars in blue (29 stars) and giant stars in red (44 stars). The latter two are stacked.
    }
    \label{MDF}
\end{figure}

Several schemes for selecting Gaia-Sausage-Enceladus members have been described in the literature, \citep[e.g.][]{Belokurov18,Helmi18,Myeong19,Feuillet21,Horta23}. Some of these selection schemes use only kinematics, whilst others also place constraints on metallicity and/or elemental abundance ratios. In particular, some of the proposed schemes include a limit on [Mg/Mn] and [Al/Fe]. However, after an inspection of the SAGA database we found that the availability of Al and Mn abundances is rather limited for our sample. We thus decided not to include any selection scheme that uses elemental abundances \citep[e.g.][]{Naidu22}. 

Table\,\ref{selections} summarises the selection schemes which we consider here. For each scheme, we use kinematics calculated using the same potential as the original authors. We select three to investigate in detail: a simple ``Sausage''-like selection, \cite{Myeong19}, and \cite{Feuillet21}. We test a very simple scheme that takes all stars with a rotational velocity typical of the halo encompassing the sausage-like overdensity, $-100 < V_{\phi} < 100$ km s$^{-1}$. This leads to a large number of stars that might belong to a high-velocity tail of the Milky Way stellar disk \citep{Nissen10,Belokurov20}.
In contrast, the scheme proposed by \citet{Myeong19} is very conservative and picks stars in a narrow range of the ``action diamond''.

As our primary selection scheme we choose that of \cite{Feuillet21}, since it has been shown to be relatively pure, while not being too restricting in regards to the number of Gaia-Sausage-Enceladus member stars \citep{Carrillo23}.
Table\,\ref{selections} also gives the number of stars selected from our full SAGA sample using each selection schemes. Using the \cite{Feuillet21} selection scheme, we have 73 stars with both Fe and Eu measurements. Of these, 70 stars have [Ba/Fe] measured, 61 stars have [Mg/Fe] measured, and 59 have both [Ba/Fe] and [Mg/Fe]. Fig.\,\ref{MDF} shows the metallicity distribution for our primary Gaia-Sausage-Enceladus sample, separated into dwarf and giant stars (see Fig.\,\ref{fullCMDHR} for separation). We note that the giant and dwarf stars do not show the same [Fe/H] distribution, at higher metallicities there are only dwarf stars. This lack of higher metallicity giant stars is likely a reflection of an inherent incompleteness of the SAGA database.

We do not include the selection schemes of \cite{Helmi18}, \cite{Naidu20}, and \cite{Horta23} in our primary elemental abundance analysis, but they are shown in App.~\ref{sect:app_trends}. \cite{Horta23} is sufficiently similar to the  scheme by \citet{Feuillet21}, while \cite{Naidu20} is similar to the ``Sausage''. Therefore, they do not give any additional insights into the properties of Gaia-Sausage-Enceladus. Although \cite{Helmi18} present the original selection scheme for the Gaia-Sausage-Enceladus, it produces a predominantly retrograde sample while more recent studies suggest that the Gaia-Sausage-Enceladus population includes stars on both retrograde and prograde orbits. The \cite{Helmi18} selection scheme could therefore potentially distort the view of the chemistry of the Gaia-Sausage-Enceladus. We find that our conclusions are not affected by our choice of selection schemes, but some selections result in trends with more scatter. 

\setlength{\extrarowheight}{5pt}

\begin{table*}
    \centering
    \caption{
    The selection schemes for Gaia-Sausage-Enceladus stars, including references. Also listed are the total number of member stars in our SAGA catalogue and the number of stars with each elemental abundance available.
    }
\label{selections}
    \begin{tabular}{cclcccc}
    \hline
    \hline
Reference & Selection criteria & Unit & \multicolumn{4}{c}{Selected number of stars}\\
 &  &  & Total & n[Eu/Fe] & n[Ba/Fe] & n[Mg/Fe]\\
\hline
  \multirow{1}{10em}{``Sausage''}   &  $-$100$< V_{t} <$100 &  km s$^{-1}$ & \multirow{1}{1em}{299} &  \multirow{1}{1em}{299}  &  \multirow{1}{1em}{286}   &  \multirow{1}{1em}{250}   \\
  \hline
    \multirow{2}{10em}{\cite{Helmi18}}    &  $-$1500$< L_{z} <$150 & kpc km s$^{-1}$ &  \multirow{2}{1em}{178}  &  \multirow{2}{1em}{178}   &  \multirow{2}{1em}{171}   &  \multirow{2}{1em}{147}   \\
            &  $E_{n} > -1.8 \times 10^{5}$ & km$^2$ s$^{-2}$ &    \\
    \hline        
   \multirow{2}{10em}{\cite{Myeong19}}    &  $-$0.07$<J_{\phi}/J_{tot}<$0.07 & &  \multirow{2}{1em}{~42} &  \multirow{2}{1em}{42}&  \multirow{2}{1em}{41}&  \multirow{2}{1em}{31} \\
            &  $(J_{z}-J_{R})/J_{tot}<$ $-$0.3 & &      \\

        \hline
    \multirow{1}{10em}{\cite{Naidu20}}   &  e >0.7 & &  \multirow{1}{1em}{264} &  \multirow{1}{1em}{264}  &  \multirow{1}{1em}{254}   &  \multirow{1}{1em}{217}   \\
    
    \hline        
  \multirow{2}{10em}{\cite{Horta23}}     &  $-$500$< L_{z} <$500 & kpc km s$^{-1}$ & \multirow{2}{1em}{~86}&  \multirow{2}{1em}{86}&  \multirow{2}{1em}{83}&  \multirow{2}{1em}{70}  \\
            &  $-1.6 \times 10^{5} <E_{n}< -1.1 \times 10^{5}$ &  km$^2$ s$^{-2}$ &    \\
    
  \hline
   \multirow{2}{10em}{\cite{Feuillet21}} &  $-$500$\leq L_{z} \leq$500 & kpc km s$^{-1}$ &  \multirow{2}{1em}{~73}&  \multirow{2}{1em}{~73}&  \multirow{2}{1em}{~70}&  \multirow{2}{1em}{~61}  \\
            &  30$\leq \sqrt{J_{r}} \leq$55 & (kpc km s$^{-1}$)$^{1/2}$ &    \\
  
\hline
\hline
    \end{tabular}
\end{table*}

\subsection{$r$-II stars}\label{sec_r2}
Stars significantly enhanced in $r$-process elements were first described by \citet{Hill02} and \citet{Sneden03}, then categorised as $r$-II stars in \cite{BC05}. 
Several definitions have been suggested, but we here use the recent definition by \cite{Holmbeck20}, which defines an $r$-II star as having $\rm [Eu/Ba]>0.0$ and $\rm[Eu/Fe]>+0.7$. In our primary sample of 73 Gaia-Sausage-Enceladus stars we have 10 $r$-II stars. 

\subsection{Uncertainties}\label{error}

\begin{figure}
    \centering
    \includegraphics[width=0.45\textwidth]{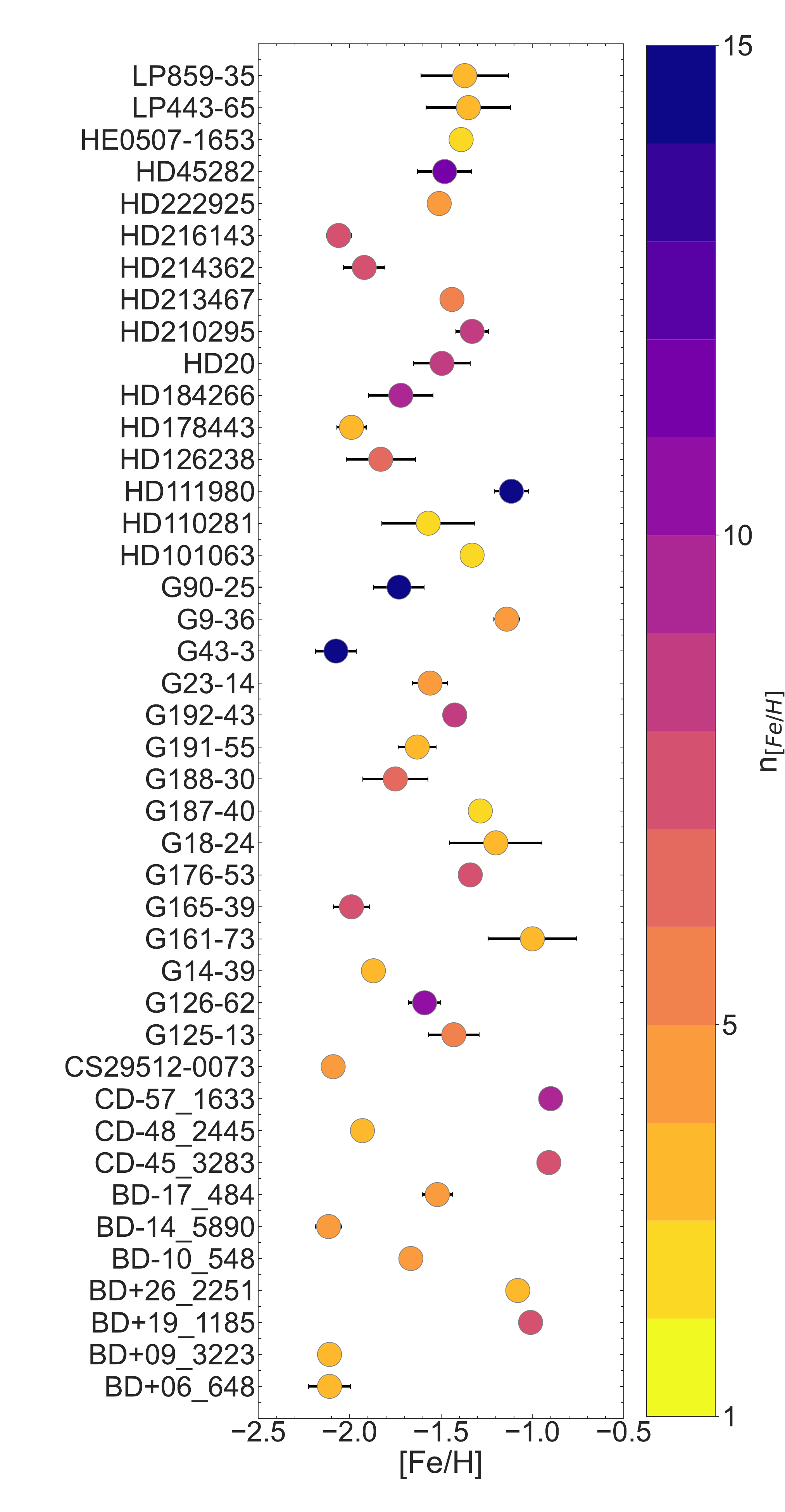}
    \caption{
    Median [Fe/H] for each star in our primary Gaia-Sausage-Enceladus sample that have two or more measurements reported in SAGA. The number of available measurements is indicated with a colour, as indicated on the colour-bar. The MAD is shown as an error-bar (see Sect.\,\ref{error}).
    }
    \label{Error-Fe}
\end{figure}

 \begin{figure}
    \centering
    \includegraphics[width=0.45\textwidth]{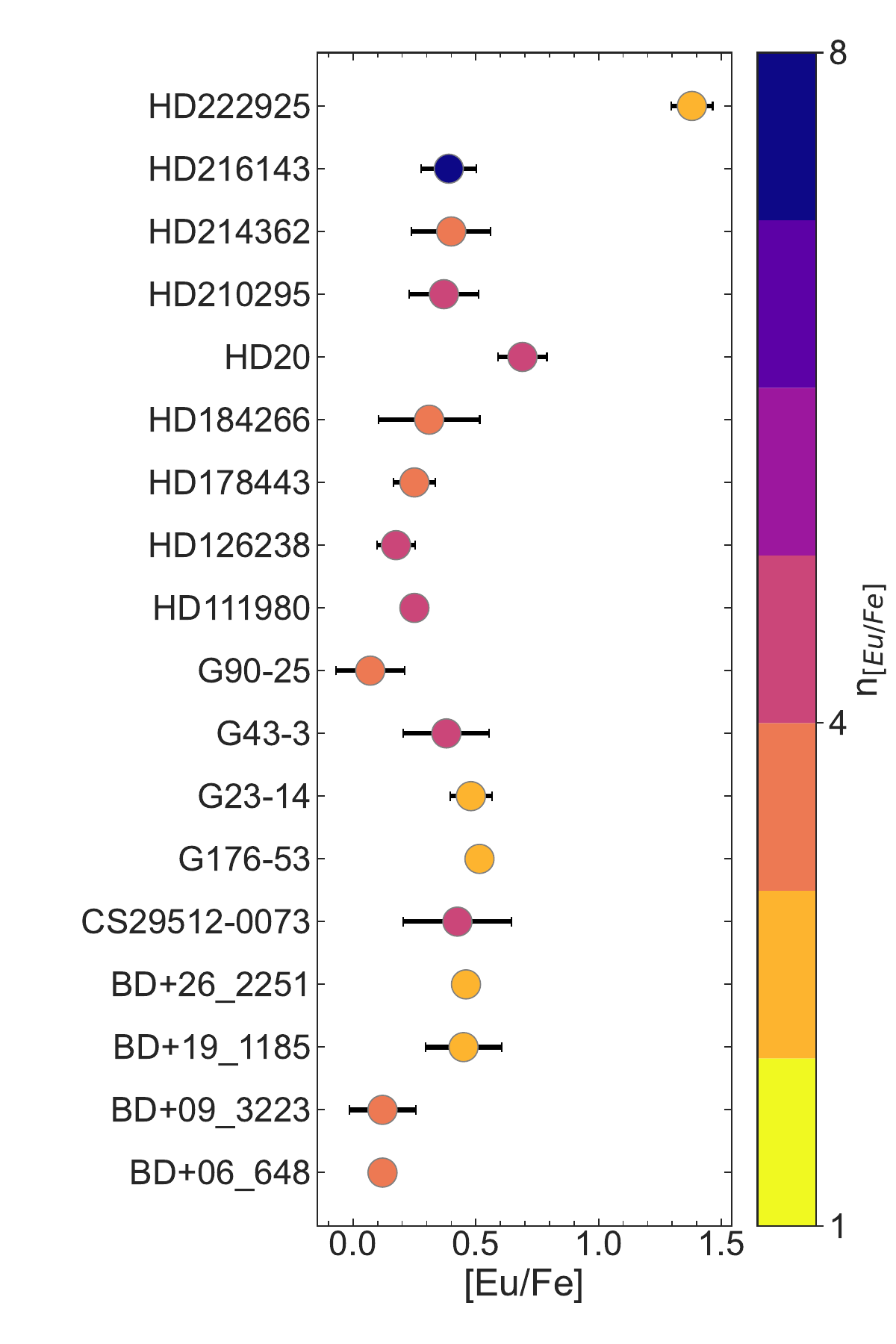}
    \caption{ Median [Eu/Fe] for stars in our primary Gaia-Sausage-Enceladus sample, which have two or more measurements reported in SAGA. The number of available measurements is indicated with a colour, as indicated on the colour-bar. The MAD is shown as an error-bar (see Sect.\,\ref{error}). }
    \label{Error-Eu}
\end{figure}

We are using a database that compiles results from different studies. Not all studies give a detailed error analysis and it is beyond the scope of this study to attempt to put all data onto the same scale and determine proper, homogeneously derived errors. Instead, we opt for a simple but reproducible scheme. For those stars with more than one entry in the SAGA database we decided to take as the final stellar parameters the median value for each parameter. The uncertainties for these values are then calculated as the Median Absolute Deviation (MAD). 

For the elemental abundance plots we will display the MAD as an error-bar. For the majority of the stars, we have only a single measurement of [Fe/H] and [Eu/Fe] (see Figs.\,\ref{Error-Fe}  and~\ref{Error-Eu}). When only one measurement is available, we show no error-bar. For elemental abundance ratios that we construct, e.g. [Eu/Ba], we calculate the error-bar by adding the MADs for the two values in quadrature. Thus we wish to alert the reader to the fact that no error-bar means a single measurement, while a substantial error-bar indicates very differing results in the literature. 

\subsection{Stellar ages}
\label{sect:ages}

We estimate ages for the dwarf stars in our sample following the methodology outlined by \citet{Sahlholdt2019} and \citet{Sahlholdt21}.
The method calculates a two-dimensional age-metallicity probability distribution function as described in \cite{Howes19}. This method is based on a Bayesian isochrone fitting methodology and produces a joint marginal likelihood $\mathcal{G}(\tau,\zeta|\textbf{x})$ of age ($\tau$) and metallicity ($\zeta$), given a set of input observables ($\textbf{x}$) in comparison with a set of model isochrones. We use a grid of isochrones from PARSEC \citep{Bressan12}, where each isochrone point is described by its initial mass ($m$), age ($\tau$), metallicity ($\zeta$), and distance modulus ($\mu$). The likelihood for each model is calculated as
\begin{equation}
\label{L}
    L(m,\tau,\zeta,\mu|\textbf{x}) = \exp \left[ -\frac{1}{2} \sum_{i} \left(\frac{x_{i}-X_{i}(m,\tau,\zeta,\mu)}{\sigma_{i}}\right)^2\right],
\end{equation}
where $X_{i}$ are the isochrone parameters at a given point, which is compared to the observable parameters of a given star $x_i$ with uncertainty $\sigma_i$. 

For each star, $\mathcal{G}(\tau,\zeta|\textbf{x})$ is the marginalisation over the mass and distance modulus, using priors $\xi(m)$ and $\psi(\mu)$, respectively, and calculated as 
\begin{equation}
\label{2dgfunc}
    \mathcal{G}(\tau,\zeta|\textbf{x}) = \int_{\mu} \int_{m} \xi(m) \psi(\mu)L(m,\tau,\zeta,\mu|\textbf{x}) \,dm \,d\mu .
\end{equation}
We use a Salpeter initial mass function \citep{Salpeter55} for $\xi(m)$ and a flat prior on distance modulus. 
As our sample is too small to consider the two-dimensional age-metallicity probability distribution, we calculate the one-dimensional $\mathcal{G}(\tau|\textbf{x})$ by marginalising over the metallicity assuming a flat prior. 

\begin{figure}
    \centering
    \includegraphics[width=0.5\textwidth]{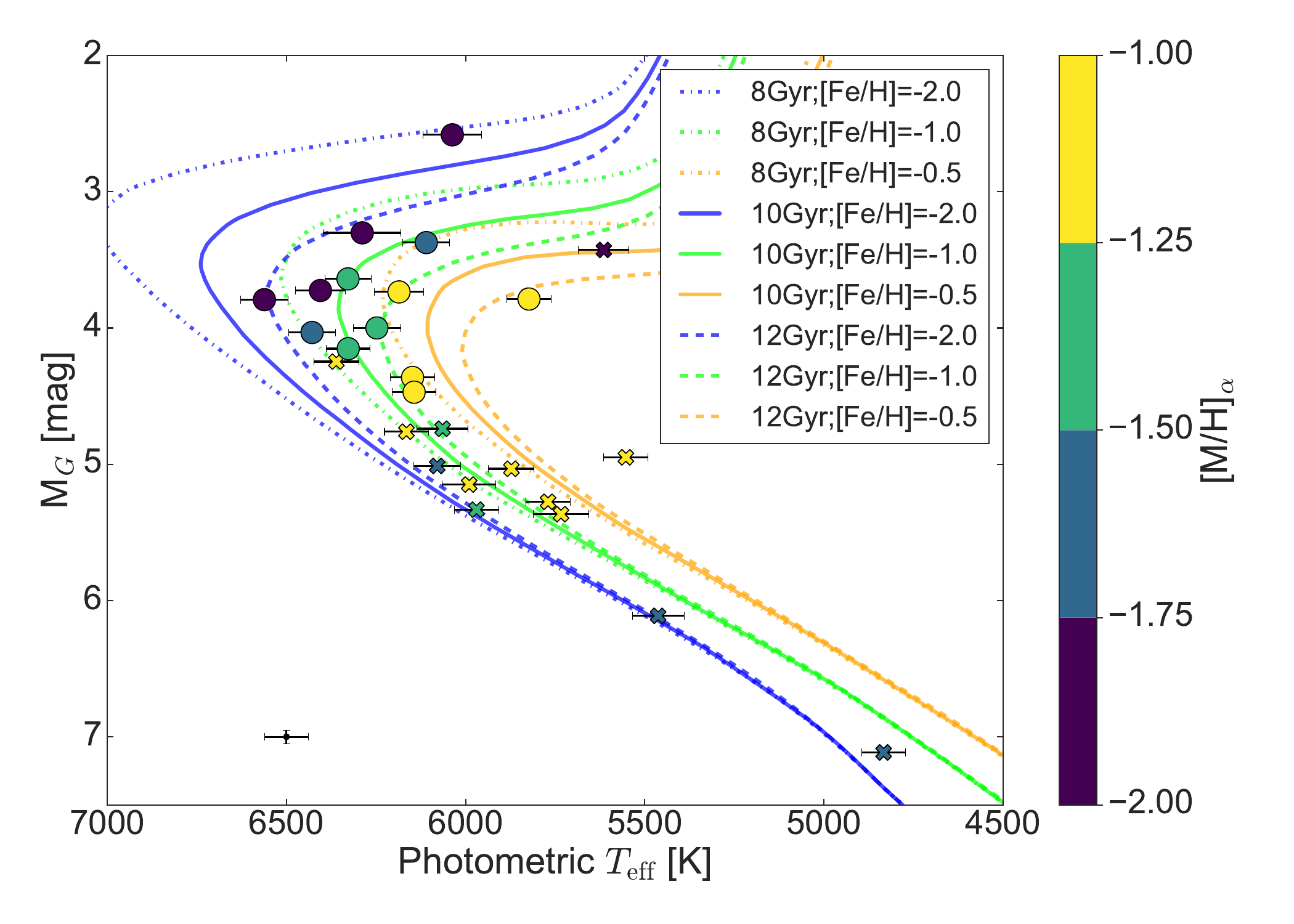}
    \caption{M$_{G}$-T$_{\rm eff}$ diagram of 26 dwarf stars with age estimates. Age input parameter uncertainties are shown with error bars, with the minimum uncertainty indicated in the bottom left corner. The circles represent the 13 stars with reliable ages and crosses represent the 13 stars with unreliable ages, both colour-coded by the adjusted metallicity ([M/H]$_\alpha$). PARSEC isochrones are plotted according to the legend. }
    \label{Mg-Teff}
\end{figure}

The observable input parameters used for each star are the $T_{\rm eff}$, apparent magnitude, parallax, and metallicity. Fig. \ref{Mg-Teff} shows the 26 dwarf stars for which we attempt to estimate ages, with the uncertainties shown by error bars. The filled circles indicate stars with a successful age estimate and cross the ones with unsuccessful age estimates. The colour indicates their adjusted metallicity ([M/H]$_\alpha$). 
In the left corner, we show the minimum uncertainties.

 The apparent magnitude is taken as the Gaia $G$-band magnitude and corrected for extinction using the reported Gaia \textsc{ag\_gspphot}. The uncertainties in Gaia~$G$ combined with the uncertainties in \textsc{ag\_gspphot} are very small, $\sim$ 0.001. Using very small uncertainties in the input parameters produces strong grid effects in our age probability distribution functions. To avoid this, we adopt a minimum uncertainty of 0.05 mag in Gaia~$G$. Adopting this minimum uncertainty has no impact on the median age or the overall shape of the probability distribution function, it only reduces the grid effects resulting in a smoother probability distribution function.

The parallax is taken from Gaia DR3 and corrected for the zero-point offset following the prescription provided by \cite{Lindegren21} with the uncertainties provided by Gaia. 

 We use the SAGA [Fe/H] and [Mg/Fe] values to calculate a metallicity that is adjusted for $\alpha$-enhancement following the equation provided by \citet{salaris93}. When estimating ages for stars with enhanced $\alpha$ abundances, an adjusted metallicity is required as ages can be over-estimated by $\pm$ 3.5~Gyr if not taken into account \citep{salaris93}.

 The $T_{\rm eff}$ values were determined from the dereddened Gaia $BP-RP$ colour and SAGA [Fe/H], using the calibration of \cite{Mucciarelli21}, reproduced below. 
\begin{equation*}
        T_{\rm eff} = \frac{5040}{\theta}
\end{equation*}
where
\begin{multline*}
\theta = b_{0} + b_{1} \cdot (BP-RP)_{0} + b_2 \cdot (BP-RP)_{0}^2 + b_{3} \cdot [Fe/H] \\
+ b_{4} \cdot [Fe/H]^2 + b_{5} \cdot [Fe/H] \cdot (BP-RP)_{0}
\end{multline*}
and
\begin{align*}
    b_{0} & = 0.4929, \quad b_{1} = 0.5092, \quad b_{2} = -0.0353, \\
    b_{3} & = 0.0192, \quad b_{4} = -0.0020, \quad b_{5} = -0.0395 
\end{align*}

The $T_{\rm eff}$ uncertainty due to the methodology is estimated to be 61~K by \cite{Mucciarelli21}. We propagate the full uncertainty accounting for the reported uncertainty from $BP$, $RP$, reddening, [Fe/H], and methodology, which has an effective minimum of 61~K. 
We use the photometric $T_{\rm eff}$ instead of the literature $T_{\rm eff}$ from SAGA because the SAGA stellar parameters are not homogeneously derived, something which is crucial for ensuring our derived stellar ages are directly comparable and on the same absolute scale. See App.\,\ref{app:teff} for more details on this choice.

\begin{figure}
    \centering
    \includegraphics[width=0.45\textwidth]{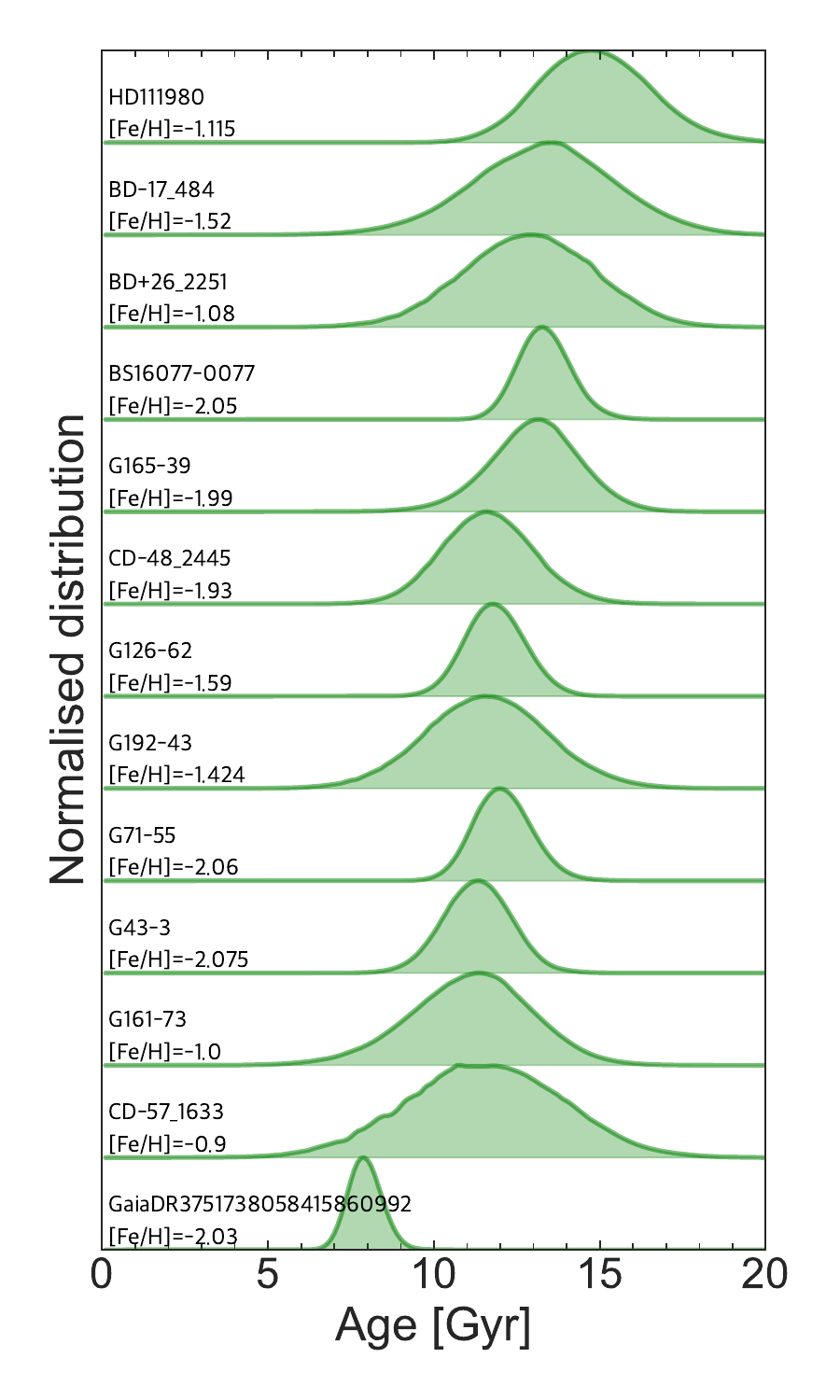}
    \caption{Age probability distribution functions for the 13 dwarf stars in our primary Gaia-Sausage-Enceladus sample with reliable ages. The remaining 13 dwarf stars have less well-behaved probability distribution functions (see App.\,\ref{app:ages}). }
    \label{GSE-ages}
\end{figure}

 Depending on the observed parameters of a star as compared to the isochrones, the derived age probability distribution function may not be smooth, Gaussian, or even singly-peaked (see \citealp{Sahlholdt2019} for a detailed discussion). We selected those stars that have well-behaved peaks as reliable age estimates. 13 out of 26 dwarf stars are selected and their age probability distribution functions are shown in Fig.\,\ref{GSE-ages}. The age probability distribution functions that did not pass our selection are shown in App.\,\ref{app:ages}. Of the 13 stars with ages, 12 have a median age older than 10 Gyr.

 We note that one star has a fairly young age, strongly constrained to be below 10 Gyr. This star, Gaia DR3 751738058415860992, has only one source in the SAGA database and the spectroscopic $T_{\rm eff}$ is 600~K cooler than the photometric $T_{\rm eff}$ used in the age analysis. In addition, the SAGA [Fe/H] (-2.03) is much lower than the Gaia \textsc{gspphot} metallicity (-1.39). 
 This young star is the dark purple circle at $T_{\rm eff}\sim6000$~K and M$_G \sim2.7$ in Fig.~\ref{Mg-Teff}. One can see from this figure that if we adopt the spectroscopic $T_{\rm eff}$, which is cooler than the photometric $T_{\rm eff}$, thereby taking the $T_{\rm eff}$ and metallicity from the same source, the star would lie along an older isochrone (the solid blue line). If the Gaia \textsc{gspphot} metallicity was adopted as well as the photometric $T_{\rm eff}$, the star would lie along a much younger isochrone, not shown in Fig.~\ref{Mg-Teff}. We have run our age analysis adopting the \textsc{gspphot} metallicity and obtained an age of 6.7 Gyr, as expected. Our computed ages using both metallicities are in agreement with the young tail of Gaia-Sausage-Enceladus found by \cite{Horta24}. In our analysis, we cross-checked our GSE sample against the young stars identified in \cite{Horta24} and found that Gaia DR3 751738058415860992 is part of their sample. This star also appears young in our analysis, providing independent verification of the existence of younger stars within the selection \cite{Feuillet21} scheme.

\section{Results -- Elemental abundances trends}
\label{sect:result}

\begin{figure*}
    \centering
    \includegraphics[width=0.95\textwidth]{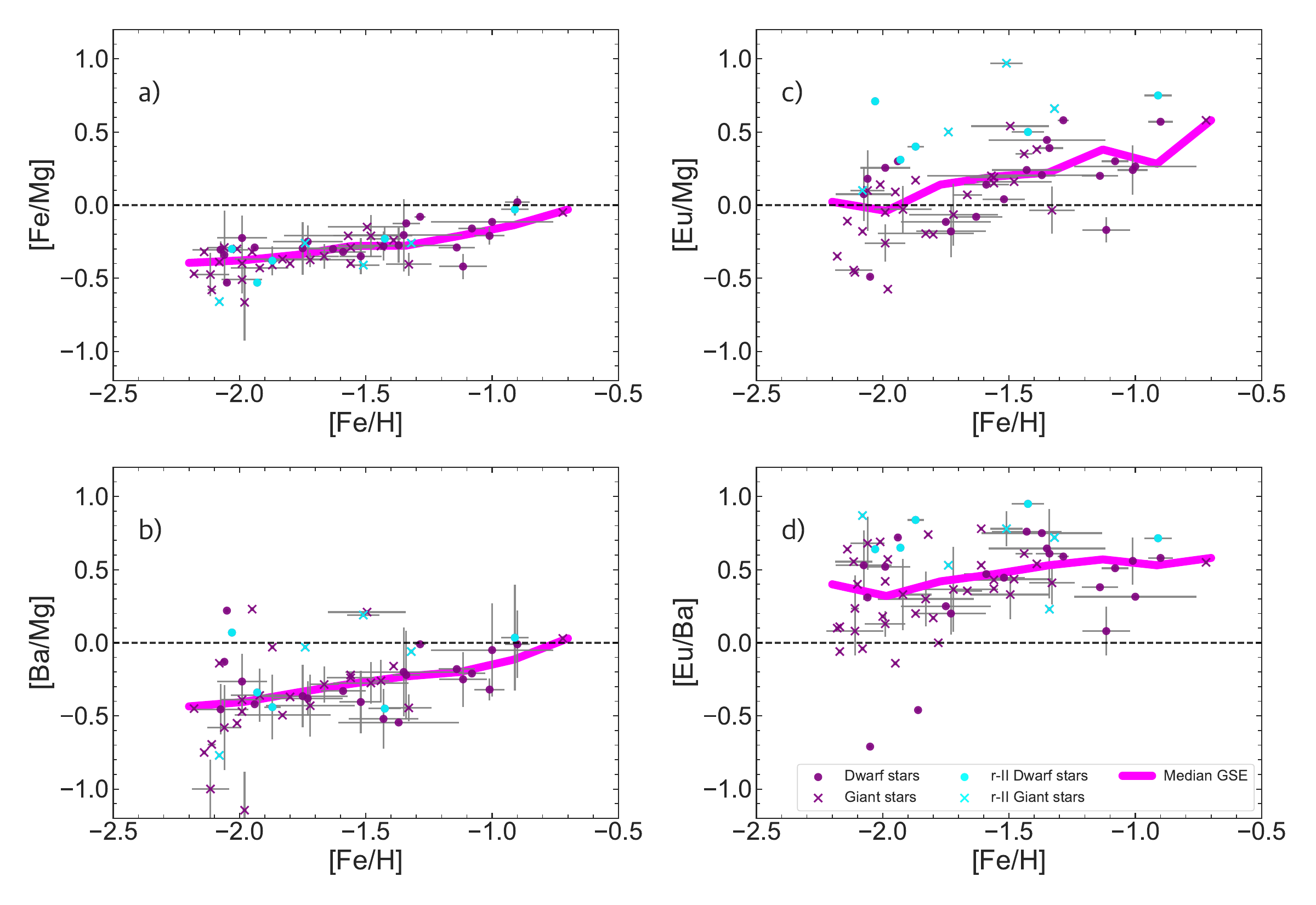}
    \caption{
    Gaia-Sausage-Enceladus stars selected from the SAGA database (Sect.\,\ref{sect:data}). Panels show elemental abundance ratios as a function of [Fe/H]: a) [Fe/Mg]; b) [Ba/Mg]; c) [Eu/Mg]; and d) [Eu/Ba]. Different symbols are dwarf (circles) and giant (x-symbols) stars. $r$-II stars are shown in cyan and $r$-normal stars shown in purple. Median trends are shown with pink lines. 
    }
    \label{Diane-trends}
\end{figure*}

\begin{figure*}
    \centering
    \includegraphics[width=1.0\textwidth]{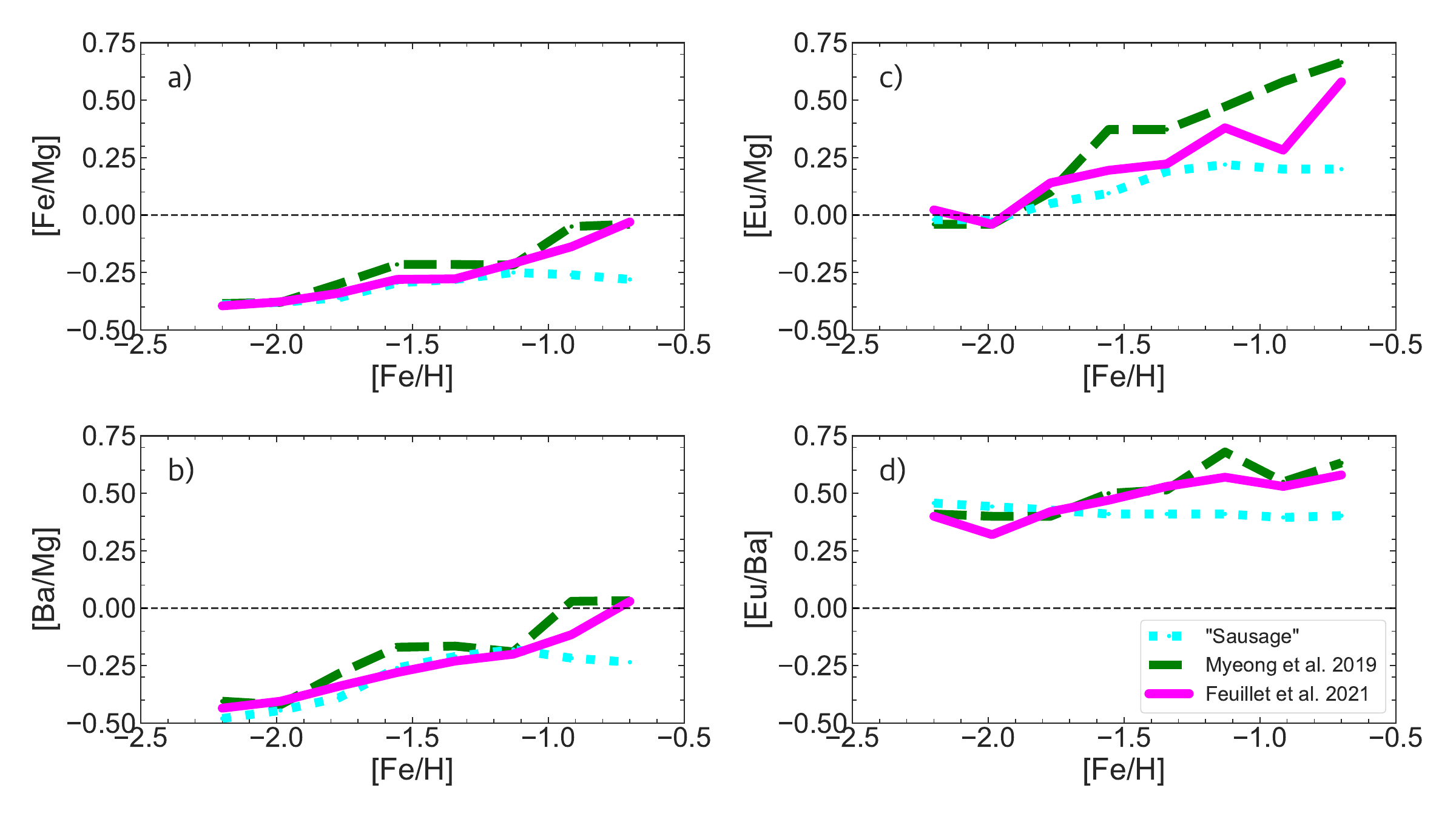}
    \caption{
    Median abundance trends for Gaia-Sausage-Enceladus stars as derived in Fig.\,\ref{Diane-trends} (pink), including the selection schemes of ``Sausage'' (cyan) and \citet[][green]{Myeong19}. Panels show trends with [Fe/H] for: a) [Fe/Mg]; b) [Ba/Mg]; c) [Eu/Mg]; and d)[Eu/Ba].
    } 
    \label{3selec-trends}
\end{figure*}

In Fig.\,\ref{Diane-trends}, we inspect four key elemental abundance ratios for the 73 stars selected using the scheme by \citet{Feuillet21}: [Fe/Mg], [Ba/Mg], [Eu/Mg], and [Eu/Ba]. We highlight both dwarf and giant stars (see Fig.\,\ref{fullCMDHR} for the division),  as well as $r$-II stars.  The trends seen in the dwarf and giant stars are not significantly different. We note that the $r$-II stars appear to follow the median trends in a similar fashion to the other stars, apart from their high Eu abundances, in accordance with their definition (Sect.\,\ref{sec_r2}). We calculate the median abundance trends, for a more straight-forward comparison between different selection schemes.
The median trend lines were calculated as the running median in 0.25\,dex bins of [Fe/H], moving in steps of 0.15\,dex. Only the data-points are taken into account, not the error-bars (for a discussion of the error-bars the reader is referred to Sect.\,\ref{error}). 

For all four abundance ratios, [Fe/Mg], [Ba/Mg], [Eu/Mg], and [Eu/Ba], we see overall rising trends with increasing [Fe/H].  However, for two of them, [Eu/Mg] and [Eu/Ba], the trends appear to be more complex and not monotonically increasing. The [Fe/Mg] distribution has the lowest scatter around the median trend line. In contrast, the abundance ratio distributions that include Eu have higher star-to-star scatter, with some stars quite far from the median trend. As Fig.\,\ref{Error-Eu} shows, [Eu/Fe] measurements from different studies can be very different from one another. As most of the stars selected from the SAGA database only have one Eu measurement in the literature, it is not possible to robustly assess if the large scatter in Fig.\,\ref{Diane-trends}\,c and d) is due to large systematic differences between studies or if the Eu abundance in the stars is truly different. A dedicated effort to obtain a set of homogeneously analysed, high-resolution spectra for Gaia-Sausage-Enceladus stars sampling the full metallicity range is needed to better quantify the intrinsic scatter. The [Ba/Mg] distribution generally has a smaller scatter than Eu, with a larger scatter at the lowest [Fe/H] end.

In order to test our decision to only analyse the elemental abundance trends of the Gaia-Sausage-Enceladus population selected following \cite{Feuillet21}, we also investigate the trends resulting from the ``Sausage'' and \cite{Myeong19} selection schemes, see Fig.\,\ref{3selec-trends}.
Overall, these trends are quite similar. However, the trends resulting from the ``Sausage'' selection scheme deviate in [Fe/Mg] and [Ba/Mg] for $\rm[Fe/H]\gtrsim-1$. In [Eu/Ba] and [Eu/Mg], similar differences appear already at $\rm[Fe/H]\approx-1.5$. These deviations at higher [Fe/H] are likely due to the inclusion of kinematically hot, disk-like Milky Way stars in the sample, which have preferentially higher [Fe/H]. 

Stars selected as Gaia-Sausage-Enceladus members by \citet{Myeong19} have very similar abundance trends to those selected by \citet{Feuillet21}, but with higher Eu abundances at $\rm[Fe/H]\gtrsim-1.5$. This may suggest that the \cite{Myeong19} selection scheme results in fewer Milky Way stars; however, because the sample selected following \cite{Feuillet21} has twice the number of stars and the resulting trends are very similar, we proceed with our analysis considering only the sample selected with the scheme by \cite{Feuillet21}. As described in Sect.~\ref{GSEselect-sect}, the selection scheme of \cite{Feuillet21} is found to be the most pure by \cite{Carrillo23} based on an analysis of a cosmological simulation of a Milky Way-like galaxy.
The references reported in the SAGA database for this sample are listed in App. \ref{sect:app_refs} in table \ref{tab:refs-diane}.

For completeness, in App.\,\ref{sect:app_trends} we show the elemental abundances and resulting elemental abundance trends for the five remaining schemes listed in Table\,\ref{selections}. These abundance trends display overall the same features as those displayed in Fig. \ref{3selec-trends}. 


\section{Literature comparison with recent elemental abundance studies}
\label{sect:literature}

Several dedicated studies have measured the abundances of heavy elements in the Gaia-Sausage-Enceladus stellar population, including \citet{Monty20}, \citet{Aguado21}, \citet{Matsuno21,Matsuno22}, and \citet{Naidu22}. 
\cite{Fishlock17} derived Eu for the low- and high-$\alpha$ halo stars from \cite{Nissen10} sample.

\cite{Aguado21} and \cite{Naidu22} derived [Eu/Ba] for 4 and 11 stars, respectively. Both studies find an elevated [Eu/Ba] in Gaia-Sausage-Enceladus, $\sim 0.7$ at [Fe/H] $\sim -1.5$ and $0.5$ at $-1.2 <$ [Fe/H] $< -0.8$, respectively, 
$\sim 0.2$ -- $0.4$ dex higher than Milky Way disk stars at the same [Fe/H] \citep{Skuladottir20}.
This is roughly consistent with our sample. 
These two studies each cover narrow but complementary [Fe/H] ranges
($-2.0 <$ [Fe/H] $< -1.4$ and $-1.2 <$ [Fe/H] $< -0.8$, respectively). Our SAGA data covers a much larger [Fe/H] range than either individual study ($-2.2 <$ [Fe/H] $< -0.7$). We note that the \cite{Aguado21} and \cite{Matsuno21} samples are included in our SAGA dataset. 

\cite{Naidu22} measured [Eu/Mg] for their stars, finding an enhancement of $\sim 0.5$, consistent with our final sample. \cite{Fishlock17} measured [Eu/Fe] to be enhanced in the \citet{Nissen10} low-$\alpha$ stars ($\rm[Eu/Fe]=0.3$ at $\rm[Fe/H]=-1.0$).

Finally, \cite{Monty20} analysed 11 stars, selected via the kinematic scheme of \cite{Myeong19}. 
The stars span $-2.5< \rm[Fe/H] <-1.0$, with one star at $-3.5$\,dex. They measured [Ba/Fe], finding that Gaia-Sausage-Enceladus is enhanced by $\sim 0.4$~dex relative to the Milky Way thick disk population. In their sample, there are no $r$-process-enhanced stars. This is inconsistent with the results from \cite{Aguado21} and our SAGA dataset. 

\section{Disentangling the star formation history of Gaia-Sausage-Enceladus}
\label{sect:discussion}

\begin{figure*}
    \centering
    \includegraphics[width=0.95\textwidth]{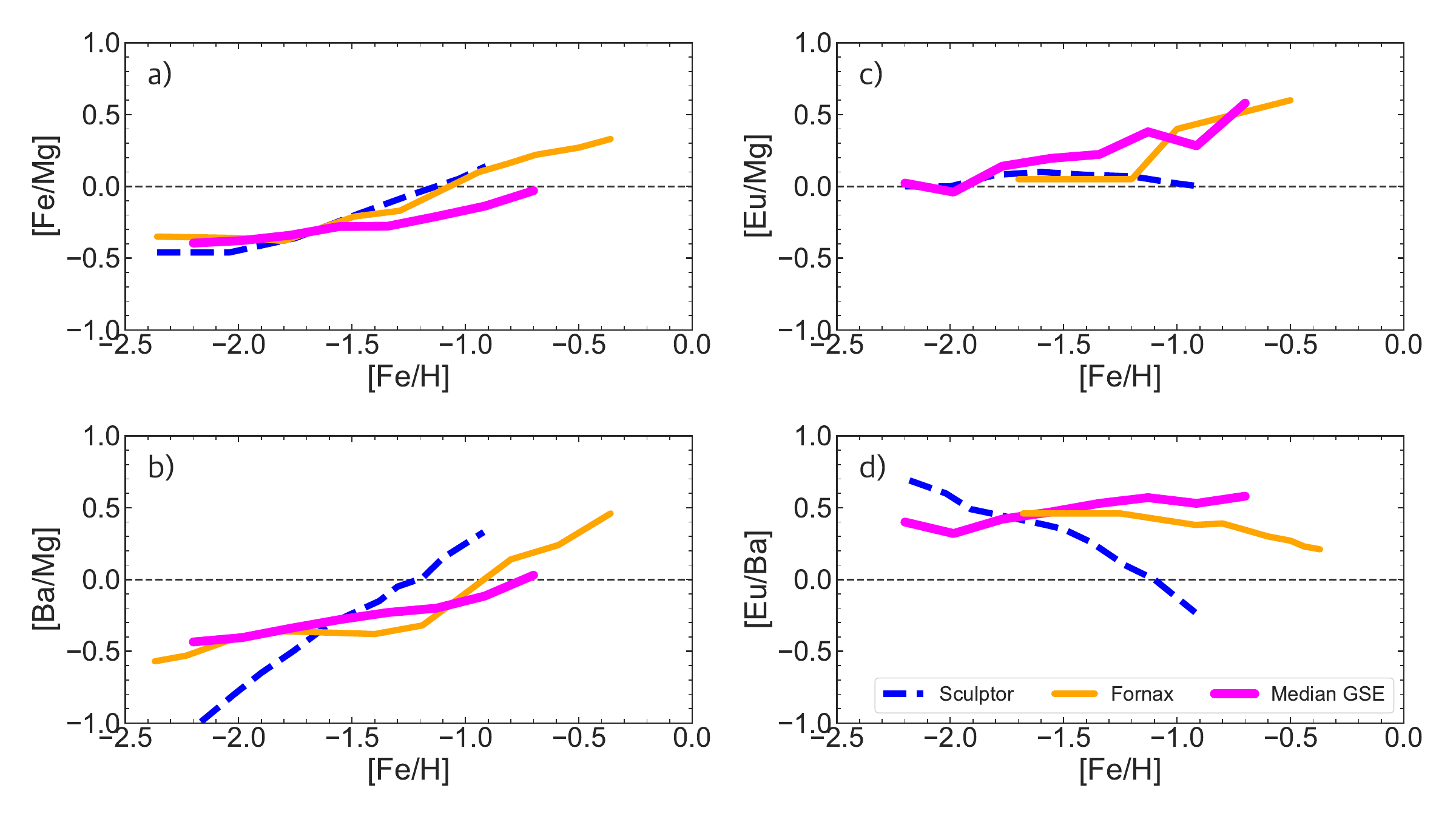}
    \caption{
    Median abundance trends of stars in the Gaia-Sausage-Enceladus (Fig.~\ref{Diane-trends}) in pink solid line, and the dSph galaxies Fornax (orange solid line) and Sculptor (blue dashed line) from \citet{Skuladottir20}. Abundance ratios are shown as a function of [Fe/H] for: a) [Fe/Mg]; b) [Ba/Mg]; c) [Eu/Mg]; and d) [Eu/Ba].
  }
    \label{Dwarfs-trends}
\end{figure*}

Our main goal is to characterise the star formation history of the Gaia-Sausage-Enceladus progenitor galaxy using constraints derived from the elemental abundance trends of its stellar debris. We assume that the debris originates from a single, ancient, low-mass galaxy. We therefore interpret the elemental abundance trends as representing an individual galaxy with its own star formation history and, consequently, its own chemical enrichment. 
\cite{Skuladottir20} use the known star formation histories of several local dwarf galaxies to place constraints on the enrichment timescale of individual nucleosynthetic channels given the observed elemental abundance trends in each galaxy's stellar population. 
We reverse the methodology of \cite{Skuladottir20} and use observed elemental abundance ratios combined with the enrichment timescales of Mg, Fe, Ba, and Eu to constrain the star formation history. 

To place the star formation history of the Gaia-Sausage-Enceladus progenitor in context, we compare the elemental abundance trends of [Fe/Mg], [Ba/Mg], [Eu/Mg], and [Eu/Ba] with the same trends measured in the nearby dSph galaxies Sculptor and Fornax (Fig.~\ref{Dwarfs-trends}), whose star formation histories are known. These two galaxies have very different star formation histories. The Sculptor dSph galaxy had a short star formation history with a strong initial starburst \citep{Bettinelli19,deBoer12} while the Fornax dSph galaxy had a more extended star formation history \citep{deBoer12}. These two extremes of star formation make them ideal objects of comparison when interpreting how the features we see in the elemental abundance trends of Gaia-Sausage-Enceladus correlate to its star formation history. 

Characterising the star formation of Gaia-Sausage-Enceladus via elemental abundance trends requires an understanding of the enrichment timescales for the individual elements. We use Mg, Fe, Ba, and Eu because these elements represent different nucleosynthetic channels, and therefore they reflect different aspects of the star formation history of a low-mass galaxy, as discussed in \cite{Skuladottir20}. 

\subsection{Nucleosynthesis of Mg, Fe, Ba, and Eu}
\label{sec:nucleo}

Core-collapse Supernovae (ccSN) are the primary end-stage of high mass stellar evolution (11 -- 40 \(\textup{M}_\odot\), \citealt{WW95}). Due to the short lifetimes of massive stars, the enrichment of elements formed primarily through ccSN can serve as a tracer of star formation \citep[e.g.][]{McWilliam13}. The element Mg is synthesised almost exclusively in ccSN \citep{WW95,WH07}. 
We use the Mg enrichment in each of the dwarf galaxy stellar populations as a direct reflection of the ongoing star formation. Fe is also synthesised in ccSN, however, this is not the only major nucleosynthetic source of Fe. 

Unlike ccSN, enrichment of elements produced in Type Ia Supernovae (SN Ia) can have a wide time delay, between $\sim$ 0.1 -- 2 Gyr \citep{Maoz14} with a typical time delay of 1 Gyr \citep{Dubay24,Palicio24}, from the epoch of star formation because the SN Ia mechanism requires an evolved binary system with a white dwarf. Fe is mainly synthesised through SN Ia \citep{WW95,Nomoto18}. The enrichment of Fe is therefore complex, depending on both immediate ccSN and delayed SNIa, however, the contribution of Fe from SNIa will be delayed compared to the contribution from ccSN.

The slow neutron capture process ($s$-process) primarily occurs in intermediate-mass Asymptotic Giant Branch (AGB) stars \citep[0.9 -- 10 \(\textup{M}_\odot\),][]{Karakas14}. Due to the range of masses, and therefore lifetimes, of these stars, the $s$-process is considered to have both a delayed and extended contribution. 
The element Ba is predominantly produced via the s-process \citep[e.g.][]{Bisterzo14}. We therefore assume that the enrichment of Ba from a single starburst will be delayed compared to the enrichment of Mg. This is because of the difference in initial mass of AGB stars and ccSN, which causes a delay of between 0.1 -- 10\,Gyr \citep{Raiteri96}, depending on the AGB star mass. 

The rapid neutron capture process ($r$-process) is less understood than the $s$-process. This process requires a large neutron flux and possible sources have existed only in theory until recently \citep{Burbidge57, Winteler12, skuladottir19, Wanajo09, Cote19, Holmbeck24}. Recent observations have confirmed neutron star mergers as a viable site for the $r$-process \citep{Abbott17, Rosswog18, Watson19}. The contribution of neutron star mergers to chemical enrichment is expected to be delayed, again due to the requirement of an evolved binary system, although the exact timescale is not currently well constrained \citep{Matteucci14, Hirai15, Rosswog18}. However, studies of stellar elemental abundance patterns suggest there is another $r$-source without significant time delay relative to ccSN \citep[e.g.][]{skuladottir19,Watson19,Cote19}.

The element Eu is produced primarily via the $r$-process and can therefore be used to constrain its timescale. \cite{Skuladottir20} demonstrated that both a quick and a delayed source of Eu is needed to explain the measured [Eu/Mg] in the Milky Way and three dSph galaxies: Sagittarius, Fornax, and Sculptor. The quick source is found to contribute Eu on timescales similar to Mg, timescales similar to ccSN, which directly traces star formation. The delayed source is found to operate on timescales longer than SN Ia. \cite{Skuladottir20} suggest that the timescales of the delayed source can be constrained by the star formation history of the Sculptor dSph galaxy. There is no evidence of an enhanced enrichment of Eu relative to Mg in Sculptor. Therefore, the delayed enrichment source of Eu did not contribute significantly during the time Sculptor was forming stars. \cite{Skuladottir20} therefore constrain the delayed source of Eu to at least $\gtrsim4$~Gyr by using the star formation history of Sculptor from \cite{deBoer12b}. However, an improved star formation history of Sculptor from \cite{Bettinelli19} finds that Sculptor only formed stars for $\sim$2~Gyr. We therefore assume that the typical timescale of the delayed $r$-process is $\gtrsim$ 2~Gyr.

\subsection{Evidence of truncated star formation in [Fe/Mg]}
\label{sec:Fe_Mg}

The elemental abundance ratio [Fe/Mg] shows the balance of the quick 
enrichment from ccSN (Mg and Fe) versus the delayed enrichment from SNIa (Fe). The low [Fe/H] stars in all three galaxies, $\rm[Fe/H]\lesssim-1.8$, 
have low [Fe/Mg] as the only source of enrichment is ccSN, see Fig.~\ref{Dwarfs-trends}a. As the SNIa begin to contribute Fe, the ratio of [Fe/Mg] increases.  
In this space, the stronger the contribution of SNIa relative to ccSN, the steeper the slope of [Fe/Mg]. As the amount of star formation decreases in a galaxy, the delayed SNIa will continue contributing Fe at a rate proportional to the star formation $\sim$1~Gyr ago, while the ccSN contribution remains proportional to the ongoing star formation. This results in a positive [Fe/Mg] at high [Fe/H] in galaxies that experience a gradual decline in star formation, as seen in Fornax and Sculptor, Fig.\,\ref{Dwarfs-trends}a. We find that the Gaia-Sausage-Enceladus trend does not extend to positive [Fe/Mg] values, suggesting that star formation was quenched and not allowed to exhaust its gas supply naturally.

\subsection{Extended star formation inferred from [Ba/Mg]}
\label{sec:Ba_Mg}

The [Ba/Mg] ratio indicates the balance between the  quick contribution of ccSN (Mg) and delayed, extended contribution of AGB stars (Ba). At low [Fe/H], this ratio reflects the strength of the initial star formation in a galaxy. This is well demonstrated in Sculptor, Fig.~\ref{Dwarfs-trends}b, which experienced a robust initial burst of star formation. Consequently, the [Ba/Mg] ratio is extremely low at low $\rm[Fe/H]\lesssim-1.8$. 
Sculptor experienced significant enrichment from ccSN before AGB stars had time to contribute. In contrast, Fornax did not have a strong initial burst of star formation, but began forming stars more gradually. This is reflected in the higher [Ba/Mg] at low [Fe/H]. The [Ba/Mg] in Gaia-Sausage-Enceladus at low [Fe/H] is very similar to Fornax, suggesting that Gaia-Sausage-Enceladus also began forming stars more gradually and did not have a strong initial burst of star formation. 

With increasing [Fe/H], the [Ba/Mg] trend reflects whether or not star formation continued for an extended period. Sculptor had a short star formation history and consequently its stellar populations shows a steep and monotonic increase in [Ba/Mg] with [Fe/H]. This reflects the increasing influence of lower mass AGB stars with time, while the ccSN contribution diminishes with the decreasing star formation.
In Fornax, the [Ba/Mg] ratio remains relatively stable until $\rm[Fe/H]\approx-1.2$, at which point it increases with a slope similar to Sculptor. The stable balance of [Ba/Mg] over a range of [Fe/H] thus reflects the extended star formation experienced by Fornax. 

In Gaia-Sausage-Enceladus stars, the [Ba/Mg] ratio increases smoothly across the entire [Fe/H] range, see Fig.~\ref{Dwarfs-trends}b. This suggests that Gaia-Sausage-Enceladus experienced an extended star formation, similar to Fornax, where AGB stars of a wider range of masses progressively contribute without a sudden decrease in the contribution from ccSN. 
However, unlike Fornax, Gaia-Sausage-Enceladus does not have a high [Fe/H] tail of increasing [Ba/Mg]. This suggests that there was no period of gradually decreasing star formation, but that Gaia-Sausage-Enceladus stopped forming stars suddenly. 

Similar to the [Fe/Mg] trends, Sculptor and Fornax extend into positive values of [Ba/Mg] due to a natural decline in star formation caused by gas loss, see Fig.~\ref{Dwarfs-trends}a,b. Gaia-Sausage-Enceladus does not reach positive values of neither [Fe/Mg] nor [Ba/Mg], suggesting an abrupt cessation of star formation in Gaia-Sausage-Enceladus.

\subsection{Star formation lasted over $2$~Gyr according to [Eu/Mg]}
\label{sec:Eu_Mg}

The [Eu/Mg] ratio reflects the balance of the quick ccSN contribution (Mg) and the multiple sources of Eu, one quick, tracking star formation and another source delayed by at least $\sim$2~Gyr (Sect.~\ref{sec:nucleo}). Thus, we assume that a flat trend of [Eu/Mg] will persist for at least the first $\sim$2~Gyr in the star formation history of a galaxy, as in seen in Sculptor, Fig.~\ref{Dwarfs-trends}c. 

Fornax shows a flat [Eu/Mg] ratio until $\rm[Fe/H]\approx -1.2$, at which point there is a increase in [Eu/Mg]. This suggests Fornax enriched its gas to $\rm[Fe/H]\approx-1.2$ after at least 2 Gyr. The subsequent increase in [Eu/Mg] is due to the onset of the delayed Eu source. 
We note that Eu measurements are unavailable in Fornax at [Fe/H] $\leq$ -1.7 dex, but $\rm[Eu/Mg]$ is measured to be $\sim 0$ at the lowest [Fe/H] available. This trend of near solar ratio of the median [Eu/Mg] is expected at lower metallicities as well. In addition, the Fornax data is sparse between $-1.3 \leq$ [Fe/H] $\leq -1.0$ \citep{Reichert20}. It is clear that at $\rm[Fe/H]< -1.0$ the [Eu/Mg] ratio in Fornax is flat and increases at higher [Fe/H], however, the transition region is unfortunately under-sampled. The sharp increase in [Eu/Mg] could actually be more gradual. 

In Gaia-Sausage-Enceladus, the [Eu/Mg] ratio starts to increase at very low [Fe/H], which suggests it was inefficient at forming stars within the first $\sim$2~Gyr, in agreement with the higher [Ba/Mg] at low [Fe/H].
However, the Gaia-Sausage-Enceladus progenitor stellar mass is estimated to be $10^{8.85}$ to $10^{9.85}$ M$_\odot$ by \citet{Feuillet20} and $4 \times 10^8$ M$_\odot$ by 
\citet{Carrillo23} using the mass-metallicity relations of \citet{Kirby13} and \citet{Ma16}. Because of the high mass of Gaia-Sausage-Enceladus relative to Fornax and Sculptor \citep[$\sim 10^7$ and $\sim 10^6$ M$_\odot$, respectively,][]{McConnachie12}, 
we would expect star formation in Gaia-Sausage-Enceladus to be more efficient than suggested by the low [Fe/H] at which the [Eu/Mg] increases. A possible mechanism for keeping the [Fe/H] low over the first $\sim$2~Gyr of star formation could be prolonged accretion of pristine gas early in its star formation phase. 
Above $\rm[Fe/H]\approx-2.0$, the [Eu/Mg] of Gaia-Sausage-Enceladus smoothly increases with [Fe/H], suggesting that as the delayed source of Eu begins to contribute, there is still enrichment of Mg from ccSN. This implies that there was no dramatic change in the star formation over the [Fe/H] range represented in Fig.~\ref{Dwarfs-trends}. 

\subsection{Quenching seen in [Eu/Ba]}\label{sec:Eu_Ba}

The [Eu/Ba] ratio represents the balance of the delayed and extended enrichment from AGB stars of a range of masses (Ba), and the two sources of $r$-process enrichment (Eu), one quick and one delayed.
In Sculptor, at low [Fe/H], the [Eu/Ba] ratio is very high and then steeply decreases across the entire [Fe/H] range. This trend is simply the inverse of the [Ba/Mg] trend because Sculptor's short star formation history does not allow for the delayed source of Eu to significantly affect the stellar elemental abundance trends. As a result, the Eu enrichment tracks closely the Mg enrichment. Therefore in Sculptor, [Eu/Ba] traces the balance between the delayed and extended enrichment of Ba from AGB stars and (only) the quick source of Eu, and reflects the strong star formation early in the history of Sculptor, which then gradually died out.

As noted above, Eu measurements are not available for Fornax at low $\rm[Fe/H]<-1.7$, however, we expect the [Eu/Ba] in Fornax at low [Fe/H] to be lower than in Sculptor and similar to that of Gaia-Sausage-Enceladus. At these [Fe/H], the timescale of Eu enrichment is the same as the timescale of Mg enrichment, therefore, the [Eu/Ba] trend is expected to be the inverse of the [Ba/Mg] trend. 

As [Fe/H] increases, [Eu/Ba] in Fornax remains relatively flat with a slight decrease. The extended star formation history of Fornax allows the delayed source of Eu to begin contributing as AGB enrichment of Ba increases. This results in a much higher [Eu/Ba] than in Sculptor.
The [Eu/Ba] ratio decreased due to the decline in star formation, resulting in an increased number of AGB contributions relative to ccSNe. The low star formation at late times in Fornax \citep{deBoer12} results in a minimal contribution of Eu from the quick source.

In Gaia-Sausage-Enceladus, at low [Fe/H], the [Eu/Ba] ratio is lower than in Sculptor, similar to [Eu/Ba] in Fornax and reflecting low initial star formation, as was noted in [Ba/Mg]. However, with increasing [Fe/H], the [Eu/Ba] ratio in Gaia-Sausage-Enceladus gradually rises. This suggests that star formation remained relatively stable throughout its star-forming period, necessitating continual contribution from the quick source of Eu to generate high amounts of Eu, while Ba production increased with time from lower mass AGB stars. Unlike in Fornax, which experienced a gradual decline in star formation, the combination of the quick and delayed sources of the $r$-process is efficient enough to increase the [Eu/Ba] ratio in Gaia-Sausage-Enceladus, suggesting it did not experience a gradual decline in star formation.

\section{Overview of the Gaia-Sausage-Enceladus\\ star formation history}

\subsection{Low initial star formation with inflows}
We find that the star formation in Gaia-Sausage-Enceladus was initially relatively low, and did not start with a strong burst as indicated by the high [Ba/Mg] and low [Eu/Ba] ratios at low [Fe/H], especially when compared to Sculptor, which did experience a strong initial burst of star formation before quickly dying out. 

The increase in [Eu/Mg] of Gaia-Sausage-Enceladus occurs at a relatively low $\rm[Fe/H]\approx-2$, suggesting that in its first 2 Gyr the progenitor galaxy enriched the gas less than expected for such a massive galaxy.
Most literature studies estimate the Gaia-Sausage-Enceladus progenitor mass to be  relatively massive compared to Fornax and Sculptor \citep[see e.g.][]{Feuillet20,  Carrillo23, McConnachie12}. 
The most likely scenario that could explain why Gaia-Sausage-Enceladus was relatively inefficient in enriching its gas, is that it was still accreting significant amounts of (near) pristine gas in the first Gyrs of star formation.

\subsection{Star formation extended and beyond 2 Gyr}
Our analysis shows that the star formation of Gaia-Sausage-Enceladus lasted for more than 2~Gyr as required by the very clear increase in [Eu/Mg] with [Fe/H]. 
Based on constraints from \citet{Skuladottir20} and an updated star formation history of Sculptor from \citet{Bettinelli19}, we assume the typical timescale of the delayed $r$-process to be $\gtrsim2$~Gyr (Sect.~\ref{sec:Eu_Mg}). Therefore, the increase in [Eu/Mg] with [Fe/H] indicates that star formation lasted longer than in Sculptor, which shows no change in [Eu/Mg].

The star formation likely remained relatively high for an extended period, supported by the high or even increasing [Eu/Ba] ratio (Sect.~\ref{sec:Eu_Ba}). Further evidence of the continued star formation is given by the shallower slope of [Fe/Mg] in Gaia-Sausage-Enceladus relative to that seen in the Sculptor and Fornax dSph galaxies.

\subsection{Quenched star formation}

Quenching of the star formation likely occurred as Gaia-Sausage-Enceladus fell into the Milky Way \citep{Gallart19}. We see evidence of quenching in our dataset from the [Fe/Mg], [Ba/Mg], and [Eu/Ba] at high [Fe/H]. 
The star formation in both Sculptor and Fornax died out naturally as the galaxies gradually lost their gas. Because of low star formation at later times, the effects of SNIa and AGB stars becomes more enhanced, resulting in supersolar values of [Fe/Mg] and [Ba/Mg] in both Sculptor and Fornax (Fig.~\ref{Dwarfs-trends}a,b). The lower values of [Fe/Mg] and [Ba/Mg] at high [Fe/H] in Gaia-Sausage-Enceladus thus suggest that its star formation was quenched, and did not die out naturally through gradual loss of gas.  

Additional evidence for quenching in Gaia-Sausage-Enceladus comes from the [Eu/Ba] ratio. In both Fornax and Sculptor, [Eu/Ba] decreases towards the end of star formation (at high [Fe/H]), but this is not seen in Gaia-Sausage-Enceladus, suggesting a lack of a tail in its star formation history.

\subsection{Literature Comparison}

Previous literature works have explored the star formation history of the Gaia-Sausage-Enceladus using different methodologies. Here we compare the main results of our analysis with those available in the literature.

\cite{Hasselquist21} uses Si and Mg elemental abundances from APOGEE DR17 to fit two different galactic chemical evolution models, flexCE \citep{Andrews17} and a model described by \citet{Lian18, Lian20}. Both models present similar characteristics: low initial star formation, increasing to a peak in star formation, leading to quenching of star formation when Gaia-Sausage-Enceladus reached $\rm [Fe/H]\approx -0.6$. The flexCE model shows that the Gaia-Sausage-Enceladus star formation peaks at $\sim$4~Gyr and quenched at $\sim$4-5~Gyr. Further evidence that quenching occurred after 2-4~Gyr is based on the deficiency in [Ce/Mg] and [(C+N)/Mg]. The model based on \citet{Lian18, Lian20} peaks earlier and is quenched relatively quickly. Therefore, we find that the model using flexCE is more consistent with our view of Gaia-Sausage-Enceladus star formation history, having a broader and more extended peak in star formation and quenching at a time of high star formation after 2-4 Gyr. 

\cite{Johnson23} use the galactic chemical evolution model VICE \citep{Johnson20} to fit the [Mg/Fe] distribution of H3 observations \citep{Conroy19}. They show that Gaia-Sausage-Enceladus had an initially low star formation that peaked within $\sim$2 Gyr and subsequently decreased until it quenched around 8 Gyr ago, suggesting that it formed stars for 5 to 6 Gyr. This model is consistent with our view of the Gaia-Sausage-Enceladus star formation history, and is also quite similar to the \citet{Hasselquist21} flexCE model, but with a slightly longer duration of star formation.

\cite{Feuillet21} derive an age distribution for Gaia-Sausage-Enceladus stars,
finding an age spread of $\sim$ 3 Gyr with a peak at $\sim$ 11
Gyr. This is consistent with our view of how long Gaia-Sausage-Enceladus
formed stars, but perhaps on the short side of our estimate.
\\

In our analysis, we limited our selection from the SAGA database to stars with $\rm -2.2 < [Fe/H] < -0.5$. From the literature, the metallicity distribution of the kinematically-selected Gaia-Sausage-Enceladus contains small numbers of stars at $\rm[Fe/H]\gtrsim-0.5$, and most of these are found to be Milky Way members using individual elemental abundances \citep[e.g.][]{Feuillet21}. In our SAGA sample we do not find any kinematically selected Gaia-Sausage-Enceladus stars with $\rm[Fe/H]>-0.7$. 
\cite{Mori24} show that the highest metallicity stars from an infalling satellite would preferentially be seen today moving with the disk of a Milky Way-like galaxy, with prograde orbits. However, a relatively low number of stars are expected to be found moving with the disk. If the Gaia-Sausage-Enceladus debris has a strong metallicity gradient in the kinematics, this could result in high [Fe/H] members missing from our sample. Any missing data at the high [Fe/H] would most significantly affect our conclusion that Gaia-Sausage-Enceladus was quenched. However, any potential internal metallicity gradient in Gaia-Sausage-Enceladus is expected to be small \citep{Khoperskov23}.

\section{Conclusions}
\label{sect:conclusion}

Our study examines the abundance trends of the $r$- and $s$-process elements, Eu and Ba, alongside Mg and Fe, across the metallicity range of Gaia-Sausage-Enceladus, $\rm-2.2<[Fe/H]<-0.5$. Utilizing a meticulously curated literature compilation from the SAGA database, we constrain the star formation of the progenitor galaxy of Gaia-Sausage-Enceladus.
Employing a unique analytical approach, we interpret elemental abundance trends, particularly focusing on the ratios [Fe/Mg], [Ba/Mg], [Eu/Mg], and [Eu/Ba] relative to [Fe/H]. Furthermore, we consider the temporal discrepancies in the nucleosynthetic processes governing these elements. By comparing our findings with known dwarf galaxies such as Fornax and Sculptor, we highlight the similarities and differences.

From these elemental abundance ratios and in comparison with the known star formation histories of Fornax and Sculptor, we place three main constraints on the star formation history of Gaia-Sausage-Enceladus.

1) At low [Fe/H], the high [Ba/Mg] and low [Eu/Ba] require that the star formation in Gaia-Sausage-Enceladus was relatively low, as compared with Sculptor where the strong initial burst of star formation drove high enrichment of Mg from ccSN before significant enrichment of Ba from AGB stars occurred, resulting in a very low [Ba/Mg].

2) The increase in [Eu/Mg] in Gaia-Sausage-Enceladus requires that star formation was extended for longer than $\sim 2$ Gyr, continuing at least until the delayed enrichment source of Eu began contributing. In contrast, star formation in Sculptor did not last long enough for the [Eu/Mg] to increase, where as star formation in Fornax was extended, resulting in high [Eu/Mg] at high [Fe/H].

3) The continued high [Eu/Ba] at high [Fe/H] in Gaia-Sausage-Enceladus requires that star formation was quenched around [Fe/H] $\sim -0.5$, preventing a period when enrichment of Ba from AGB stars dominated over enrichment of Eu. This would have resulted in a decrease in [Eu/Ba], as is seen Fornax and expected in galaxies that naturally exhaust their gas supply. 
Additionally, the maximum [Fe/Mg] and [Ba/Mg] of $\sim 0$ in Gaia-Sausage-Enceladus, as compared to the super-solar values in Fornax and Sculptor, imply that Gaia-Sausage-Enceladus did not naturally consume its gas supply.

Looking ahead, we anticipate that the forthcoming 4MOST data will provide valuable insights into the Milky Way accreted populations understanding, particularly for this major accretion event, Gaia-Sausage-Enceladus. Surveys which include data from the halo, disk, and bulge \citep[Surveys 1-4][]{HelmiS1,ChristliebS2,ChiappiniS3,BensbyS4}, as well as information from dwarf galaxies and their stellar streams in Survey\,14 \citep{skuladottirS13}, will greatly improve our understanding of how the Milky Way formed and evolved with greater accuracy than ever before.

\begin{acknowledgements}
We thank Haining Li for sharing the Gaia DR3 IDs for their targets \citep{Li22}. We thank the referee for the useful comments that improved the paper.

H.E., S.F. and D.F. were supported by a project grant from the Knut and
Alice Wallenberg Foundation (KAW 2020.0061 Galactic Time
Machine, PI Feltzing). This project was supported by funds from the Crafoord foundation (reference 20230890).
D.F.~acknowledges funding from the Swedish Research Council grant 2022-03274.
\'{A}.S.~acknowledges funding from the European Research Council (ERC) under the European Union’s Horizon 2020 research and innovation programme (grant agreement No. 101117455).
\end{acknowledgements}

%
%

\bibliographystyle{aa}
\bibliography{GSE}

\begin{appendix} 
\section{ Ba-rich stars not considered in the analysis}\label{app:Bastars}

To ensure the meaningfulness of the elemental abundance trends, we have chosen to exclude Ba-rich stars. These stars do not provide insights into a galaxy's chemical enrichment but are the results of the evolution in binary systems. Apart from being high in Ba, such stars can also have enhanced carbon abundances, due to a mass transfer from an Asymptotic Giant Branch companion. We, therefore, checked the C abundances in all stars and found that all the stars listed in Table~\ref{Ba_stars} exhibit both high Ba and high C abundances, and are therefore removed from our sample. 

\begin{table}
    \centering
    \caption{ Ba-rich stars selected as Gaia-Sausage-Enceladus members, following \cite{Feuillet21} selection, that were not considered in the analysis.}
    \begin{tabular}{ccc}
    \hline
    \hline
  Gaia ID & [Fe/H] & [Ba/Fe] \\
    \hline
  Gaia DR3 2982933097213087616  & --1.39 & 1.85 \\
  Gaia DR3 3686190458143012352  & --2.04 & 2.40 \\
  Gaia DR3 2722263350106417152  & --1.20 & 1.17 \\
  Gaia DR3 2601354871056014336  & --2.09 & 1.55 \\
  Gaia DR3 2448568453248165504  & --2.11 & 2.20 \\
    \hline
    \hline
    \end{tabular}
    \label{Ba_stars}
\end{table}

\section{SAGA database literature references for Gaia-Sausage-Enceladus members}\label{sect:app_refs}

Table \ref{tab:refs-diane} gives the literature references from which the stellar atmospheric parameters and elemental abundances are gathered by the SAGA database for the Gaia-Sausage-Enceladus members following the \citet{Feuillet21} selection scheme. These are the stars used in the elemental abundance analysis and interpretation of the star formation history.

\begin{table*}  
    \centering
     \caption{SAGA database literature references used to obtain the median stellar parameters and abundances for each star selected using the \cite{Feuillet21} selection scheme.}
\resizebox{\textwidth}{!}{%
\begin{tabular}{ll}
\hline
Object & Reference \\
\hline
2MASSJ00453930-7457294 &  \cite{Hansen18}  \\
2MASSJ02412152-1825376 &  \cite{Hansen18}  \\
2MASSJ03084611-4747083 &  \cite{Jacobson15}  \\
2MASSJ05254724-3049100 &  \cite{Sakari18}  \\
2MASSJ13164824-2743351 &  \cite{Hansen18}  \\
2MASSJ14043762+0011117 &  \cite{Sakari18}  \\
2MASSJ17273886-5133298 &  \cite{Hansen18}  \\
2MASSJ19202070-6627202 &  \cite{Hansen18}  \\
BD+06\_648 &  \cite{Wako08}, \cite{Melendez01}, \cite{Burris00}, \cite{Aoki17}  \\
BD+09\_3223 &  \cite{Johnson02b}, \cite{Johnson01}, \cite{Carney03}, \cite{Johnson02a}, \cite{Burris00}  \\
BD+19\_1185 &  \cite{Boesgaard11}, \cite{Roederer14}, \cite{Roederer10}, \cite{Simmerer04}  \\
BD+26\_2251 &  \cite{Fulbright00}, \cite{Smiljanic09}  \\
BD-10\_548 &  \cite{Ishigaki12}, \cite{Ishigaki13}, \cite{Ishigaki10}, \cite{Carney03}  \\
BD-14\_5890 &  \cite{Ishigaki12}, \cite{Ishigaki13}, \cite{Charbonnel05}, \cite{Ishigaki10}, \cite{Carney03}  \\
BD-17\_484 &  \cite{Boesgaard11}, \cite{Ishigaki12}, \cite{Ishigaki13}, \cite{Ishigaki10}  \\
BD-18\_271 &  \cite{Ishigaki12}, \cite{Ishigaki13}, \cite{Ishigaki10}, \cite{Carney03}, \cite{Melendez02}  \\
BS16077-0077 &  \cite{Allen12}  \\
BS16080-0175 &  \cite{Allen12}  \\
CD-45\_3283 &  \cite{Nissen10}, \cite{Gratton03}, \cite{Caffau05}, \cite{Hansen12}, \cite{Smiljanic09}, \cite{Yan16}, \cite{Nissen11}  \\
CD-48\_2445 &  \cite{Hansen17}, \cite{Asplund06}, \cite{Melendez10}  \\
CD-57\_1633 &  \cite{Nissen10}, \cite{Gratton03}, \cite{Hansen12}, \cite{Nissen02}, \cite{Tan09}, \cite{Caffau05}, \cite{Smiljanic09}, \cite{Yan16}, \cite{Nissen11}  \\
CS29512-0073 &  \cite{Roederer14}, \cite{Johnson07}, \cite{Masseron12}, \cite{Hansen12}, \cite{Allen12}  \\
G125-13 &  \cite{Nissen10}, \cite{Ishigaki12}, \cite{Ishigaki13}, \cite{Ishigaki10}, \cite{Nissen11}  \\
G126-62 &  \cite{Nissen07}, \cite{Akerman04}, \cite{Nissen02}, \cite{Roederer14}, \cite{Roederer10}, \cite{Simmerer04}, \cite{Fabbian09}, \cite{Asplund06}, \cite{Amarsi19}, \cite{Nissen04}  \\
G14-39 &  \cite{Ishigaki12}, \cite{Ishigaki13}, \cite{Ishigaki10}  \\
G161-73 &  \cite{Nissen10}, \cite{Roederer14}, \cite{Nissen11}  \\
G165-39 &  \cite{Ishigaki12}, \cite{Ishigaki13}, \cite{Shi07}, \cite{Ishigaki10}, \cite{Gehren06}, \cite{Boesgaard05}, \cite{Stephens02}  \\
G176-53 &  \cite{Nissen10}, \cite{Ishigaki12}, \cite{Ishigaki13}, \cite{Roederer10}, \cite{Simmerer04}, \cite{Yan16}, \cite{Nissen11}  \\
G18-24 &  \cite{Ishigaki12}, \cite{Ishigaki13}, \cite{Ishigaki10}  \\
G187-40 &  \cite{Ishigaki12}, \cite{Ishigaki13}  \\
G188-30 &  \cite{Roederer14}, \cite{Ishigaki12}, \cite{Ishigaki13}, \cite{Ishigaki10}, \cite{Boesgaard05}, \cite{Stephens02}  \\
G191-55 &  \cite{Boesgaard11}, \cite{Roederer10}, \cite{Simmerer04}  \\
G192-43 &  \cite{Ryan01}, \cite{Nissen10}, \cite{Boesgaard11}, \cite{Charbonnel05}, \cite{Roederer10}, \cite{Simmerer04}, \cite{Melendez10}, \cite{Nissen11}  \\
G23-14 &  \cite{Ishigaki12}, \cite{Ishigaki13}, \cite{Roederer10}, \cite{Simmerer04}  \\
G43-3 &  \cite{Ishigaki12}, \cite{Ishigaki13}  \\
G71-55 &  \cite{Roederer14}, \cite{Sakari18}  \\
G88-23 &  \cite{Roederer14}  \\
G9-36 &  \cite{Roederer10}, \cite{Simmerer04}, \cite{Boesgaard05}, \cite{Stephens02}  \\
G90-25 &  \cite{Roederer14}, \cite{Roederer10}, \cite{Simmerer04}  \\
Gaia DR3 4821671436294995456 &  \cite{Aguado21}  \\
Gaia DR3 6183013242623029504 &  \cite{Aguado21}  \\
Gaia DR3 751738058415860992 &  \cite{Li22}  \\
HD101063 &  \cite{Roederer10}, \cite{Simmerer04}  \\
HD110281 &  \cite{Li13}, \cite{Carney03}  \\
HD111980 &  \cite{Nissen10}, \cite{Ishigaki12}, \cite{Gratton03}, \cite{Reddy06}, \cite{Ishigaki10}, \cite{Hansen12}, \cite{Nissen02}, \cite{Fulbright00}, \cite{Ishigaki13}, \cite{Tan09}, \cite{Smiljanic09}, \cite{Yan16}, \cite{Nissen11}  \\
HD126238 &  \cite{Roederer14}, \cite{Gratton00}, \cite{Sneden98}, \cite{Roederer12}, \cite{Hansen12}  \\
HD178443 &  \cite{McWilliam95II}, \cite{Roederer14}, \cite{McWilliam95I}  \\
HD184266 &  \cite{Ishigaki12}, \cite{Gratton00}, \cite{Fulbright03}, \cite{Takada-Hidai02}, \cite{Roederer14}, \cite{Ishigaki13}, \cite{Roederer10}, \cite{Simmerer04}, \cite{Carney03}  \\
HD20 &  \cite{Barklem05}, \cite{Zhang11}, \cite{Gratton00}, \cite{Fulbright03}, \cite{Carney03}, \cite{Christlieb04}, \cite{Burris00}, \cite{Ren12}  \\
HD210295 &  \cite{Alves-Brito10}, \cite{Ishigaki12}, \cite{Fulbright00}, \cite{Ishigaki13}, \cite{Fulbright03}, \cite{Roederer10}, \cite{Simmerer04}  \\
HD213467 &  \cite{Jonsell05}, \cite{Ishigaki12}, \cite{Gratton00}, \cite{Ishigaki13}, \cite{Carney03}  \\
HD214362 &  \cite{Ishigaki12}, \cite{Ishigaki10}, \cite{Burris00}, \cite{Roederer14}, \cite{Ishigaki13}, \cite{Roederer10}, \cite{Simmerer04}, \cite{Carney03}  \\
HD216143 &  \cite{Wako08}, \cite{Fulbright03}, \cite{Saito09}, \cite{Mishenina02}, \cite{Melendez01}, \cite{Burris00}, \cite{Mishenina17}, \cite{Mishenina01}, \cite{Fulbright00}, \cite{Johnson02b}, \cite{Johnson01}, \cite{Johnson02a}  \\
HD222925 &  \cite{Roederer21}, \cite{Roederer18}, \cite{DenHartog21}, \cite{Gratton00}, \cite{Roederer20}  \\
HD224959 &  \cite{Karinkuzhi21}, \cite{Masseron10}, \cite{VanEck03}  \\
HD45282 &  \cite{Gratton00}, \cite{Shi07}, \cite{Fulbright03}, \cite{Saito09}, \cite{Mishenina02}, \cite{Mishenina01}, \cite{Roederer14}, \cite{Fulbright00}, \cite{Carney03}, \cite{GarciaPerez06-O}, \cite{GarciaPerez06-LiBe}  \\
HE0507-1653 &  \cite{Karinkuzhi21}, \cite{Aoki07}  \\
HE1305+0007 &  \cite{Goswami06}, \cite{Beers07}  \\
KIC10737052 &  \cite{Matsuno21halo}  \\
KIC12253381 &  \cite{Matsuno21halo}  \\
KIC5953450 &  \cite{Matsuno21halo}  \\
KIC6611219 &  \cite{Matsuno21halo}  \\
LP443-65 &  \cite{Ishigaki12}, \cite{Ishigaki13}, \cite{Ishigaki10}  \\
LP859-35 &  \cite{Ishigaki12}, \cite{Ishigaki13}, \cite{Ishigaki10}  \\
RAVEJ015656.3-140211 &  \cite{Sakari18}  \\
RAVEJ040618.2-030525 &  \cite{Rasmussen20}  \\
RAVEJ044208.2-342114 &  \cite{Rasmussen20}  \\
RAVEJ051727.4-134235 &  \cite{Sakari18}  \\
RAVEJ074824.3-483141 &  \cite{Rasmussen20}  \\
RAVEJ094634.8-062653 &  \cite{Sakari18}  \\
RAVEJ130200.0-084328 &  \cite{Sakari18}  \\
RAVEJ161228.4-054142 &  \cite{Sakari18}  \\
SMSSJ080050.54-724620.6 &  \cite{Jacobson15}  \\
\hline
\end{tabular}}
    \label{tab:refs-diane}
\end{table*}

\section{Elemental abundance trends for different selection schemes}
\label{sect:app_trends}

For completeness, we show the elemental abundances and their derived median trends for the other five Gaia-Sausage-Enceladus selection schemes listed in Table\,\ref{selections}.
The error-bars represent the MAD for stars with multiple measurements of a particular elemental abundance ratio. These error-bars are not taken into account when deriving the the trends. We also include the $r$-II stars and high-light them. They follow the general trends apart from that their Eu abundances are high.

\begin{figure*}
    \centering
    \includegraphics[width=\textwidth]{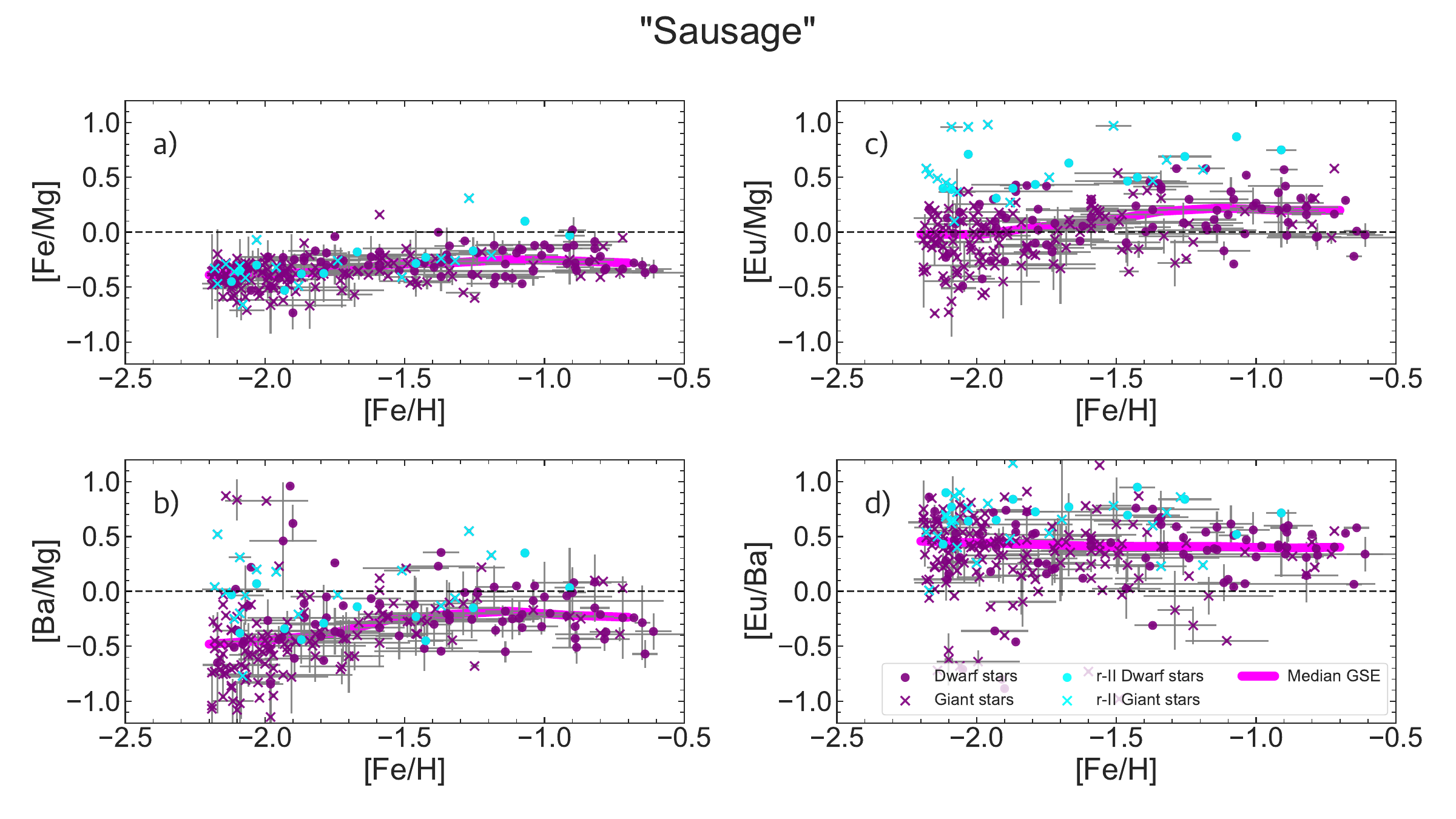}
    \caption{ The Gaia-Sausage-Enceladus stars in the SAGA data sample selected according to the ``Sausage'' scheme. Different types of stars are identified according to the legend. The running median is also shown.
}
    \label{Xnselections1}
\end{figure*}

\begin{figure*}
    \centering
    \includegraphics[width=\textwidth]{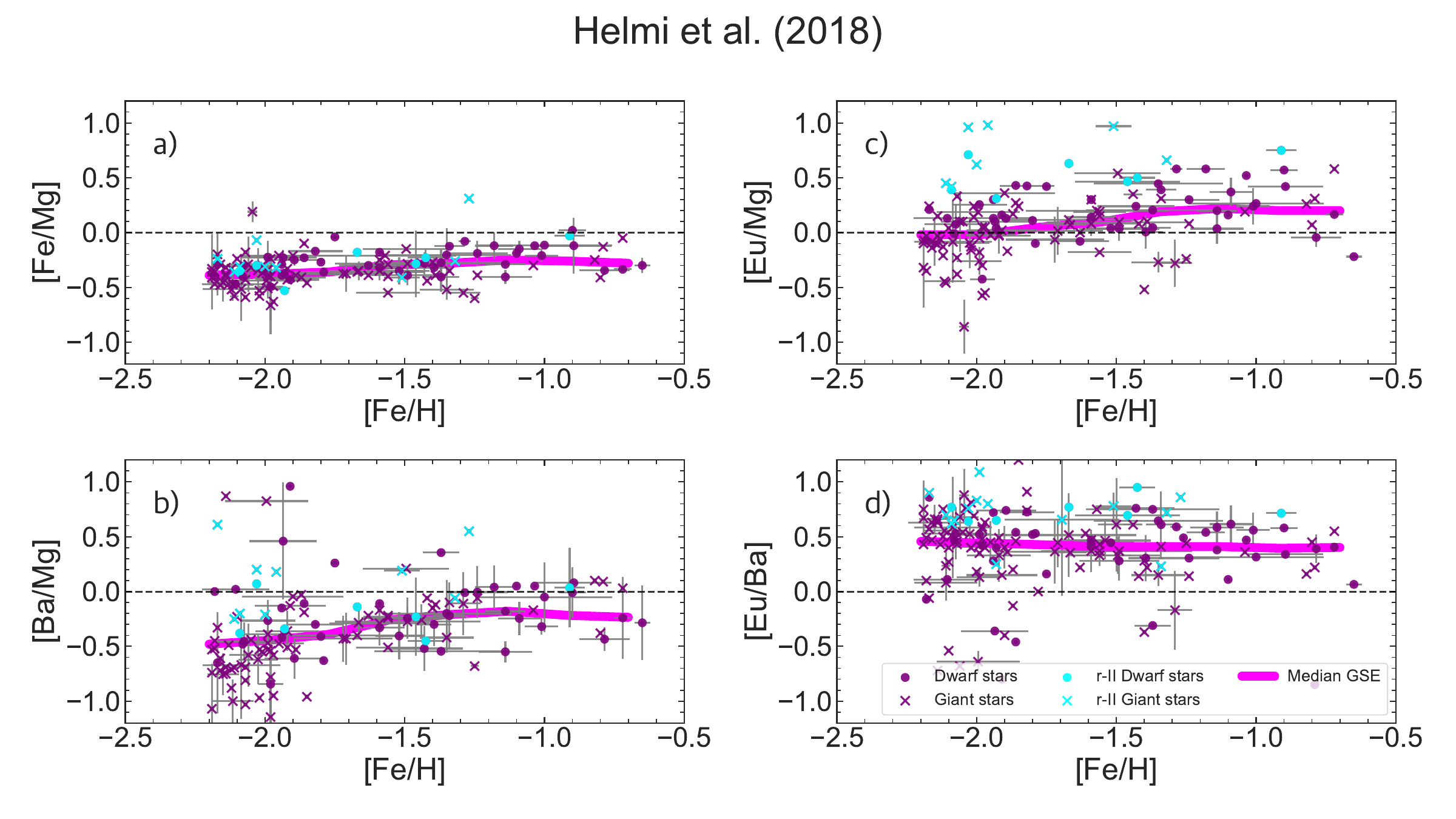}
    \caption{ The Gaia-Sausage-Enceladus stars in the SAGA data sample selected according to the scheme from \cite{Helmi18}. Different types of stars are identified according to the legend. The running median is also shown.
    }
    \label{Xnselections2}
\end{figure*}

\begin{figure*}
    \centering
    \includegraphics[width=\textwidth]{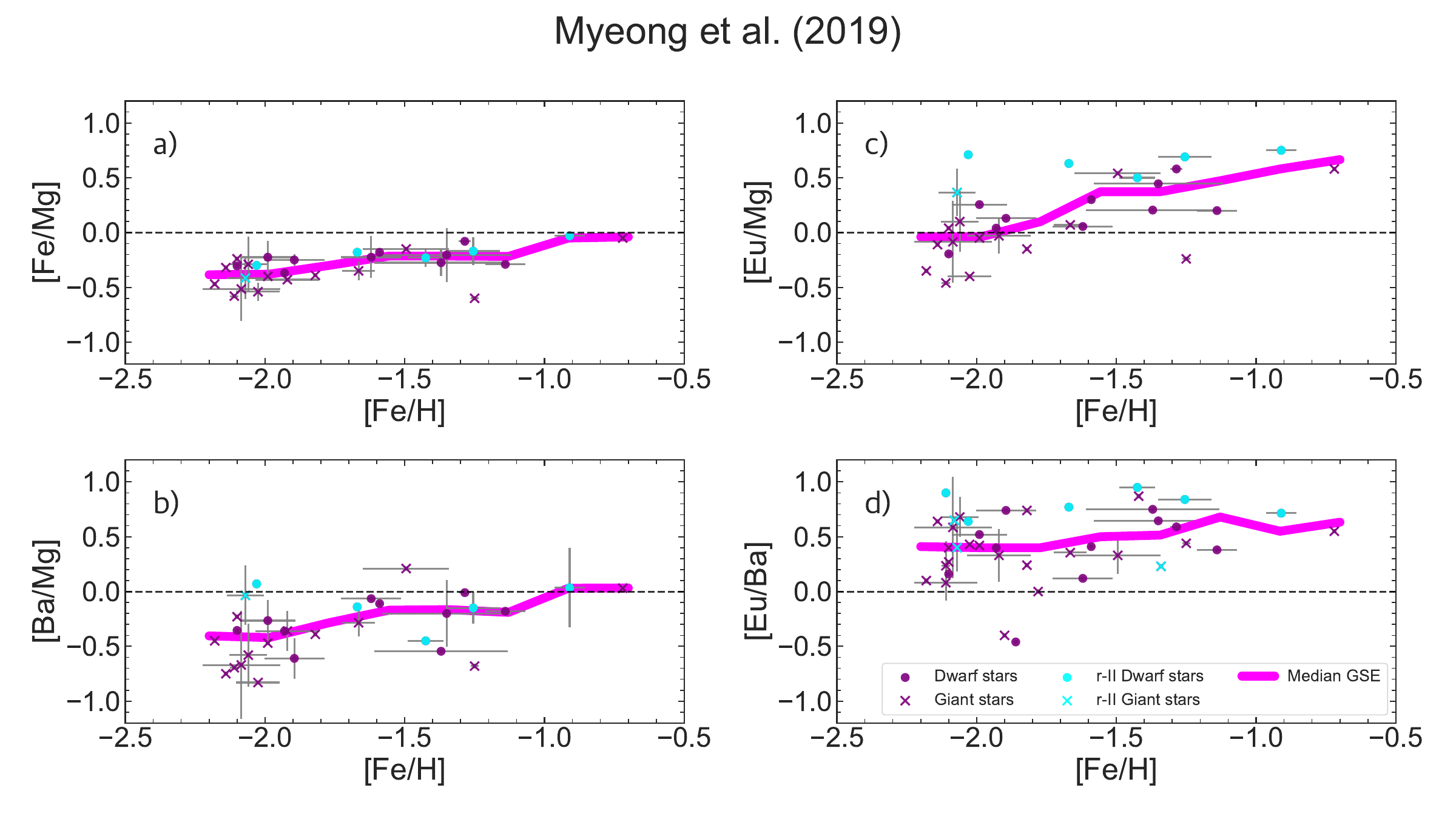}
    \caption{The Gaia-Sausage-Enceladus stars in the SAGA data sample selected according to the scheme from \cite{Myeong19}. Different types of stars are identified according to the legend. The running median is also shown.
    }
    \label{Xnselections3}
\end{figure*}

\begin{figure*}
    \centering
    \includegraphics[width=\textwidth]{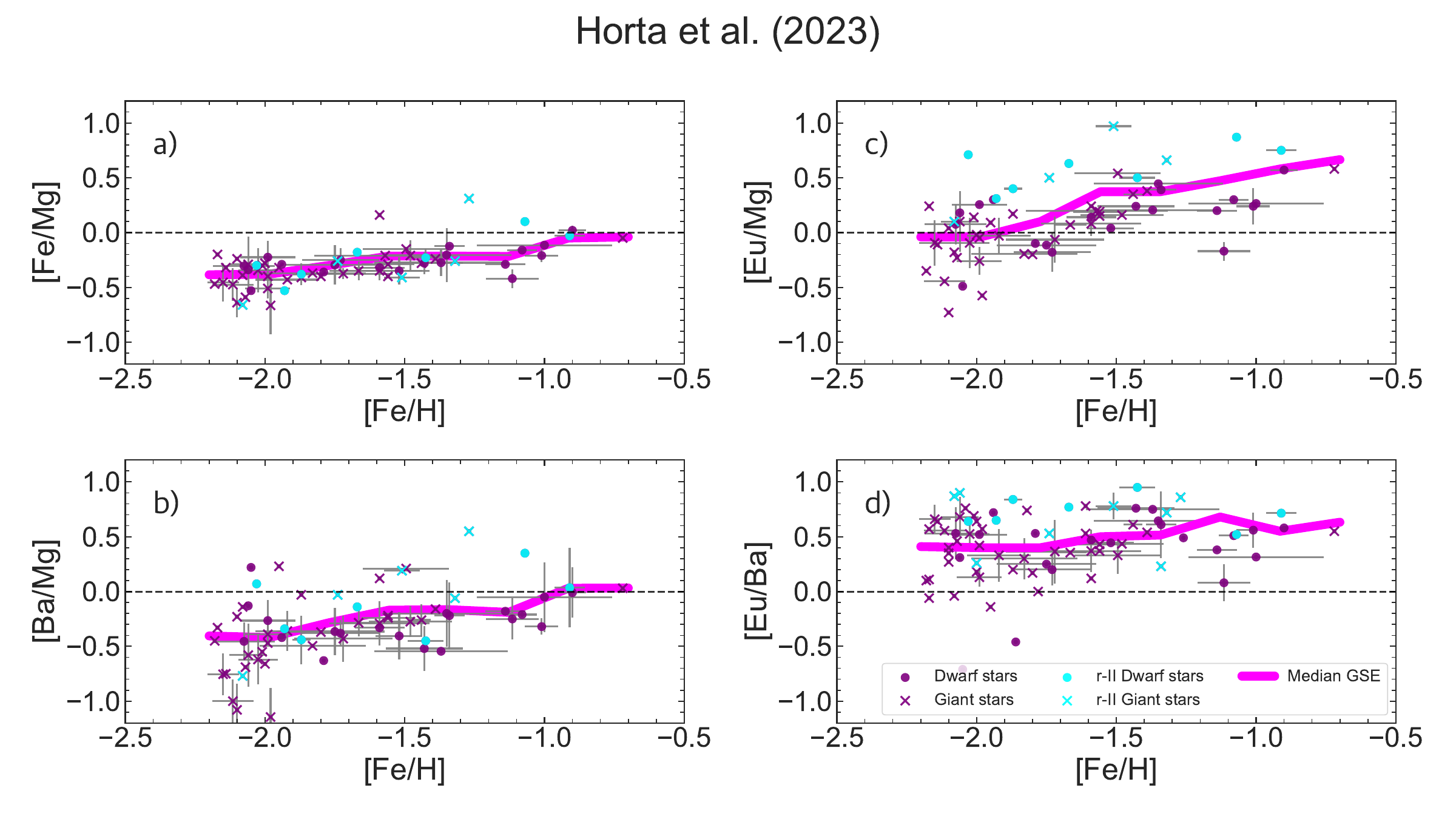}
    \caption{The Gaia-Sausage-Enceladus stars in the SAGA data sample selected according to the scheme from \cite{Horta23}. Different types of stars are identified according to the legend. The running median is also shown.
    }
    \label{Xnselections4}
\end{figure*}

\begin{figure*}
    \centering
    \includegraphics[width=\textwidth]{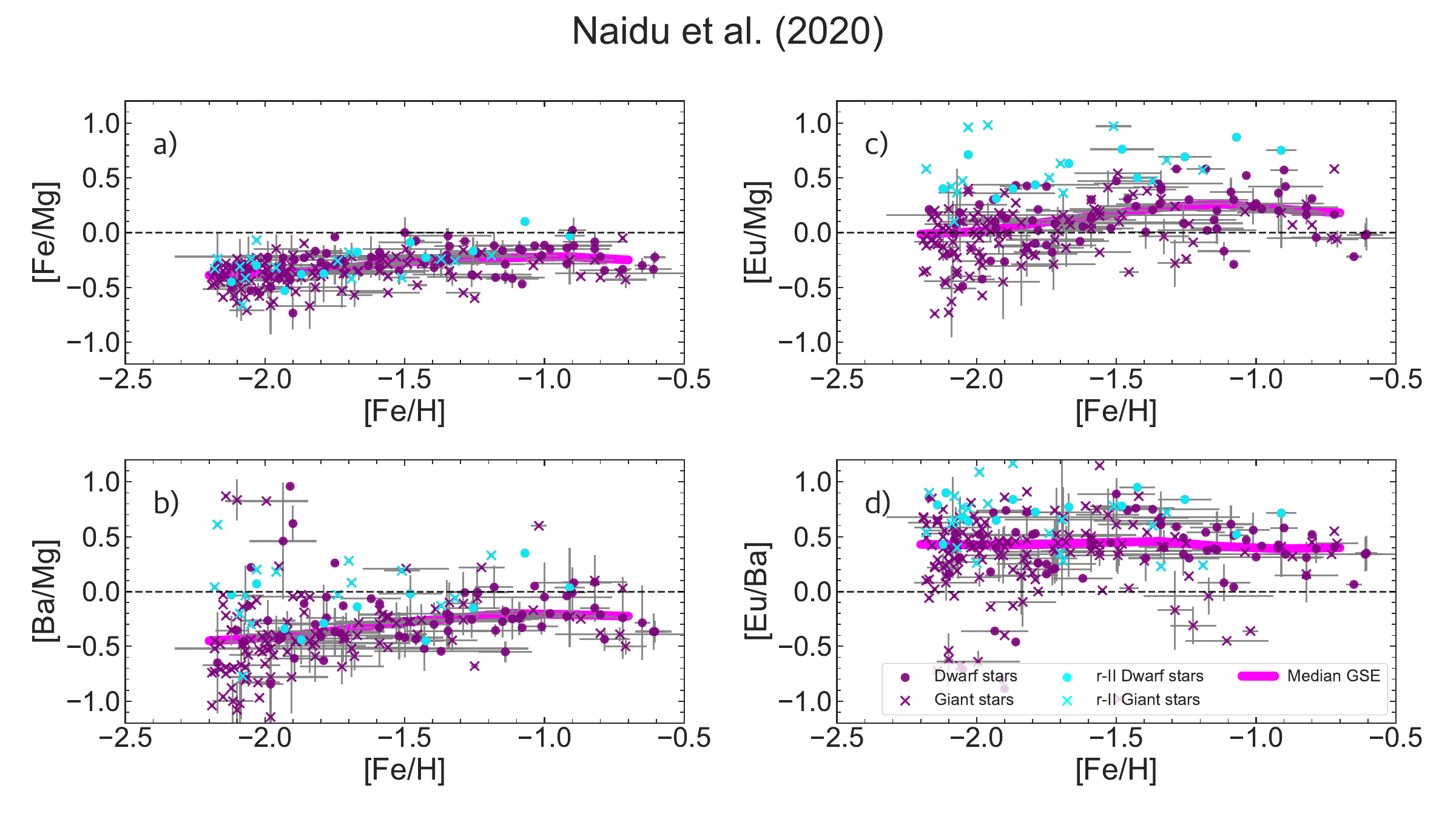}
    \caption{The Gaia-Sausage-Enceladus stars in the SAGA data sample selected according to the scheme from \cite{Naidu20}. Different types of stars are identified according to the legend. The running median is also shown. 
    }
    \label{Xnselections5}
\end{figure*}

\section{Effective temperatures and stellar ages}

\subsection{Effective temperatures -- additional checks}
\label{app:teff}

\begin{figure}
    \centering
    \includegraphics[width=0.5\textwidth]{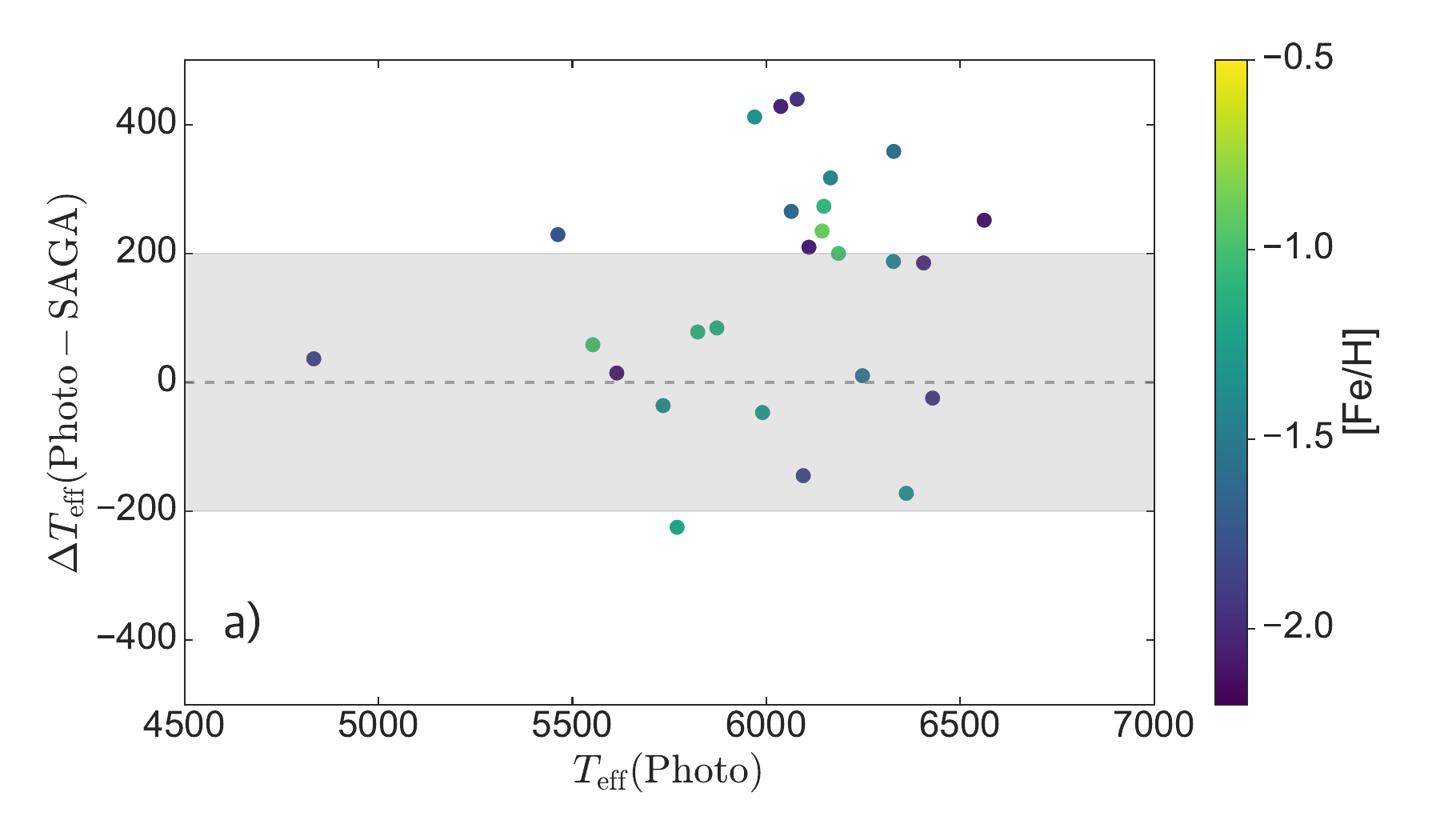}
    \includegraphics[width=0.5\textwidth]{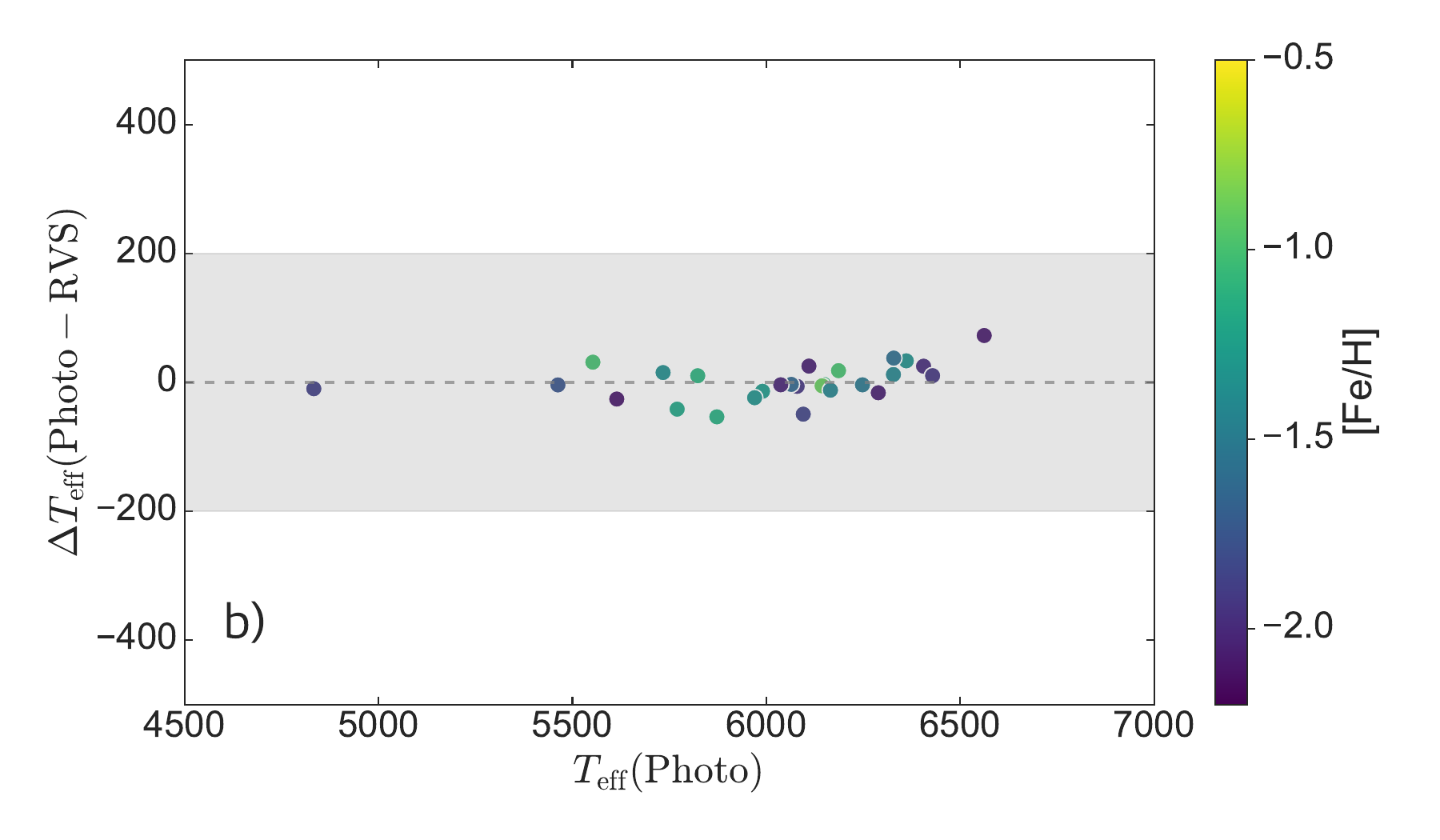}
    \caption{a) The difference between photometric $T_{\rm eff}$ and median $T_{\rm eff}$ in the SAGA database as a function of photometric $T_{\rm eff}$. b) The difference between photometric $T_{\rm eff}$ and Gaia RVS $T_{\rm eff}$ as a function of photometric $T_{\rm eff}$. In both panels, the colour-coding indicates the median SAGA [Fe/H].}
    \label{app:figTeffphoto}
\end{figure}

When estimating the stellar ages (Sec.~\ref{sect:ages}), we opted to use the photometric $T_{\rm eff}$ instead of the SAGA $T_{\rm eff}$. 
As mentioned above, the SAGA catalogue is a literature compilation, therefore stellar parameters are not homogeneously derived and some stars have $T_{\rm eff}$ reported from more than one study. This is not ideal for precise comparisons of stellar ages on a common absolute scale. In Fig.\,\ref{app:figTeffphoto} we compare the photometric $T_{\rm eff}$ calculated in Sec.~\ref{sect:ages} with the median SAGA literature $T_{\rm eff}$. The standard deviation is 211 K and the mean difference is 161 K, suggesting an offset in the median SAGA literature $T_{\rm eff}$  with our photometric $T_{\rm eff}$.

One could also consider using the Gaia RVS $T_{\rm eff}$. We compare Gaia RVS $T_{\rm eff}$ with the photometric $T_{\rm eff}$, Fig.\,\ref{app:figTeffphoto} b, we find the standard deviation is 27 K and the mean difference is 0.5 K. However, we decided to adopt the photometric $T_{\rm eff}$ for our Gaia-Sausage-Enceladus sample.

\subsection{Age estimates for 13 dwarf stars with poorly constrained probability distribution functions}
\label{app:ages}

\begin{figure}
    \centering
    \includegraphics[width=0.5\textwidth]{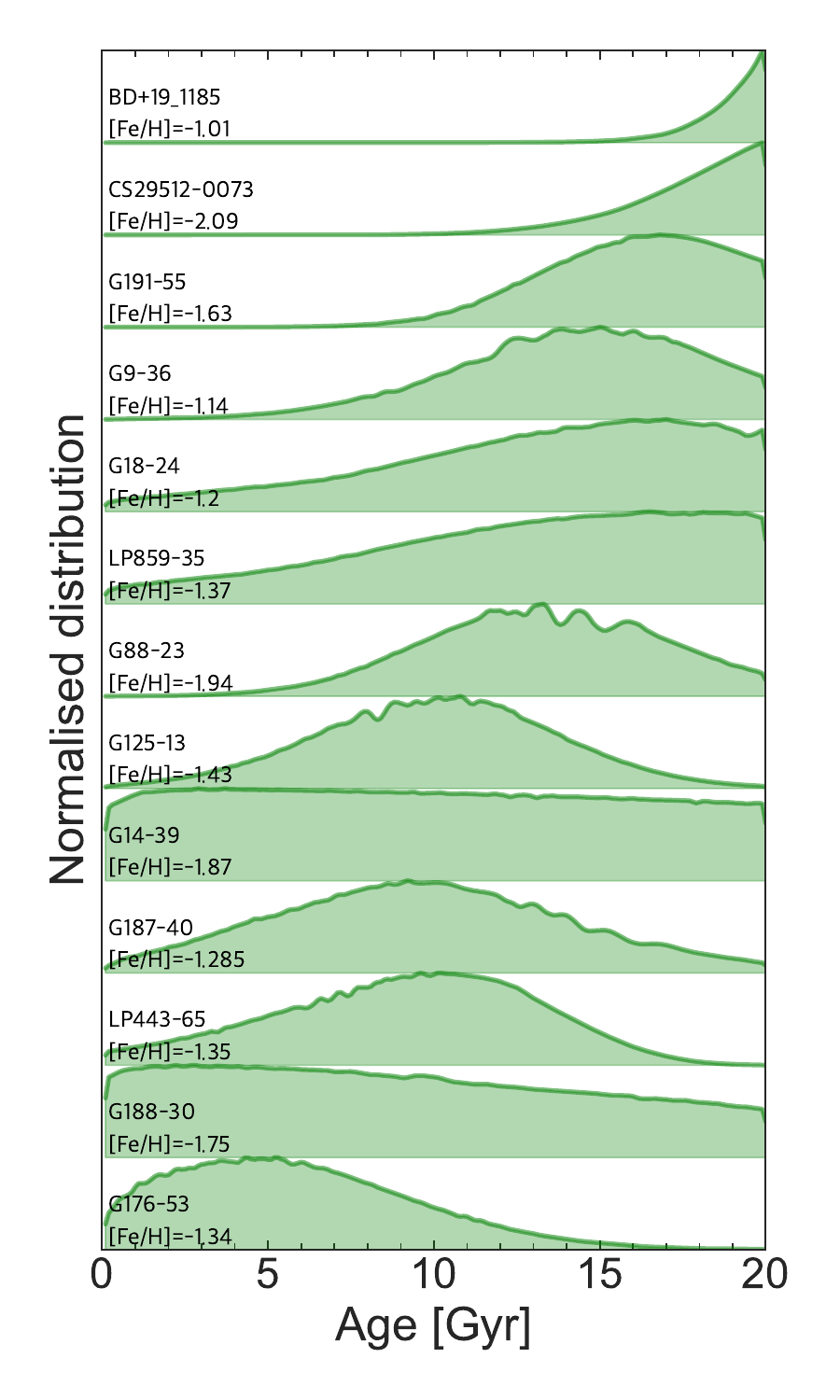}
    \caption{Age probability distribution functions for 13 dwarf stars selected as members of Gaia-Sausage-Enceladus according to \cite{Feuillet21} scheme that do not meet our quality criteria. }
    \label{GSE-dw-Ages}
\end{figure}

For completeness, we show in Fig.\,\ref{GSE-dw-Ages} the probability distribution functions for the 13 dwarf stars, that do not meet our quality criteria, described in Sec~\ref{sect:ages}. No age estimates are listed for these stars in Table\,\ref{tab:data}.

\end{appendix}

\end{document}